\documentclass[11pt,a4paper]{article}
\usepackage{xcolor}
\usepackage{ifpdf}
\usepackage{jheppub}
\usepackage{rotating}
\usepackage[bb=boondox]{mathalfa}
\usepackage{comment}

\usepackage{tikz-cd}
\usepackage{xcolor}
\usepackage{soul}

\usepackage{graphicx} 
\usepackage{mathrsfs}
\usepackage{amsmath,amsfonts,amssymb}
\usepackage{mathtools}
\usepackage{amsmath,amsfonts,amssymb}
\usepackage{graphicx}
\usepackage{dcolumn}
\usepackage{bm}
\usepackage{hyperref}
\usepackage[mathlines]{lineno}
\usepackage{float}
\usepackage{blindtext}
\usepackage{titlesec}
\title{Sections and Chapters}
\usepackage[toc,page]{appendix}
\usepackage{xcolor}
\usepackage{graphicx}
\usepackage[normalem]{ulem} 
\graphicspath{ {./images/} }
\usepackage[thinc]{esdiff}
\usepackage[euler]{textgreek}
\usepackage{braket}
\usepackage{physics}
\usepackage{xfrac}
\usepackage{soul}
\usepackage{stmaryrd}
\usepackage{pifont}
\usepackage{subfigure}
\usepackage{caption}
\usepackage{lipsum}
\usepackage{appendix}

\allowdisplaybreaks

\title{Gravitational Wave Signatures of \texorpdfstring{$\mathrm{U}(1)_X$}{U(1)X} Breaking and Right-Handed Neutrino Dynamics}

\author[a,b]{Arnab Chaudhuri,}
\author[c]{Priya Mishra,}
\author[c]{and Rukmani Mohanta }
\affiliation[a]{Division of Science, National Astronomical Observatory of Japan, Mitaka, Tokyo 181-8588, Japan.}
\affiliation[b]{Department of Physics, School of Advanced Science, Vellore Institute of Technology, Vellore-632014, Tamil Nadu, India.}
\affiliation[c]{School of Physics,  University of Hyderabad, Hyderabad - 500046,  India.}
\emailAdd{arnab.chaudhuri@nao.ac.jp}
\emailAdd{mishpriya99@gmail.com}
\emailAdd{rmsp@uohyd.ac.in}

\abstract{
The Standard Model (SM) leaves several fundamental questions unanswered, including the origin of neutrino masses, the baryon asymmetry of the Universe, and the nature of dark matter. Motivated by these gaps, we investigate an extension of the SM with an additional local $U(1)_X$ gauge symmetry and a complex scalar singlet that spontaneously breaks this symmetry via its vacuum expectation value. The extended framework naturally accommodates three right-handed neutrinos (RHNs) to ensure anomaly cancellation and implements a type-I seesaw mechanism for active neutrino masses. We utilized Casas-Ibarra parameterization to systematically reconstruct the Yukawa coupling matrix which automatically satisfy the observed neutrino data. Furthermore, we estimate the key parameters of the first-order phase transition and compute the resulting stochastic gravitational wave spectrum, demonstrating that it can lie within the reach of forthcoming experiments such as LISA, DECIGO, BBO, and the Einstein Telescope. The right-handed neutrinos also open a viable path for thermal leptogenesis, providing a unified link between neutrino mass generation, baryogenesis, and gravitational wave signatures. Our results demonstrate that this minimal $U(1)_X$ scenario remains a promising probe for physics beyond the Standard Model, accessible through upcoming gravitational wave and neutrino experiments.
}

\DeclareUnicodeCharacter{2212}{\ensuremath{-}}
\begin{document}
\maketitle 
\flushbottom

\preprint{}
\section{Introduction}
The Standard Model (SM) has long held the crown of success for its remarkable explanation of fundamental particles and their interactions. Despite its triumphs, the SM falls short in several crucial areas. It cannot explain why the Universe is matter-dominated, nor does it shed light on the nature of dark matter or dark energy. The presence of massive neutrinos and the unexplained flavor and mass structure of leptons are additional gaps, all point to the limitations of the SM in providing a complete picture of fundamental physics. Therefore,  in order to address these challenges, it is quite natural to extend the SM, either by expanding its symmetry $SU(3)_C\, \times SU(2)_L\, \times U(1)_Y$ structure or by introducing new particles. Such frameworks are part of the physics beyond the standard model~(PBSM).

Although rooted in general relativity, gravitational waves~(GW) may also be a promising probe for PBSM. The groundbreaking detection of GW events by the LIGO observatory~\cite{LIGOScientific:2016aoc} in 2015, has opened an entirely new window into the early Universe. These events may carry the traces of early Universe dynamics, like, cosmological first-order phase transitions~(FOPT), which are far beyond the reach of SM as per Lattice study~\cite{Kajantie:1996mn}. The NANOGrav experiment~\cite{NANOGrav:2020bcs} has recently reported signs of a stochastic gravitational wave background, which have been interpreted as possible signatures of a FOPT~\cite{Addazi:2020zcj, Nakai:2020oit, Li:2020cjj, Ratzinger:2020koh, Chaudhuri:2025ybh}. Lattice Monte-Carlo simulation shows that, within the SM framework, the electroweak phase transition~(EWPT) could be a FOPT for Higgs mass below $\sim 75$~GeV, but for the observed 125~GeV Higgs, it proceeds as a smooth crossover~\cite{Kajantie:1996mn}. The underlying reason is the presence of cubic term ($\phi^3$) in the finite-temperature corrections to the scalar potential, which decides the height of potential barrier between the symmetric (false vacuum) and broken (true vacuum) phase. For a Higgs mass of 125~GeV, the cubic term becomes almost negligible, reducing the potential barrier and leading to crossover.

There are several approaches to enhance the potential barrier in model building perspective. The most popular way is to introduce scalar singlet to the SM, that couple strongly with Higgs~\cite{Profumo:2007wc, Vaskonen:2016yiu, Beniwal:2017eik, Banta:2022rwg, Fu:2023nrn, Srivastava:2025oer}, in contrast to the Ref.~\cite{Okada:2018xdh,  Brdar:2018num, Okada:2020vvb}, where FOPT is realized in negligible coupling of Higgs with  scalar singlet. Another alternative is to incorporate a two Higgs-doublet model, which supply additional light bosonic degrees of freedom~\cite{Su:2020pjw, Banta:2022rwg,Dorsch:2014qja, Chaudhuri:2024vrd}. There are models based on new Higgs multiplets~\cite{Manohar:2006ga, Ham:2010ha, Bandyopadhyay:2021ipw}, additional $U(1)$ extension of SM~\cite{Okada:2018xdh, Hasegawa:2019amx, Okada:2020vvb}, minimal supersymmetric extensions ~\cite{Athron:2019teq, Zhao:2021dxa,Haba:2019qol}, frameworks with additional fermions~\cite{Gabelmann:2021ohf, Banta:2022rwg}, SMEFT based approaches~\cite{Camargo-Molina:2024sde,Qin:2024dfp}, and scenario involving varying Yukawa couplings~\cite{Baldes:2016rqn}, etc. The GW generated from such FOPTs are termed as stochastic GWs due to the presence of several GW sources that emerge during the tunneling of the Universe from symmetric to a broken phase through bubbles formation~\cite{Caprini:2018mtu, Mazumdar:2018dfl, Caprini:2019egz, Hindmarsh:2020hop}. Driven by the energy difference between inside and outside of bubble walls (latent heat), expanding bubbles collide and produce GWs~\cite{Turner:1990rc, Kosowsky:1991ua, Kosowsky:1992rz, Turner:1992tz, Kosowsky:1992vn}. Furthermore, the motion of expanding bubbles stirs the surrounding plasma, inducing turbulence~\cite{Kamionkowski:1993fg, Kosowsky:2001xp, Dolgov:2002ra, Gogoberidze:2007an, Caprini:2009yp} and generating sound (pressure) waves~\cite{Hindmarsh:2013xza, Hindmarsh:2015qta, Hindmarsh:2016lnk}, both of which act as additional sources of GWs. Among all of them, sound waves produce the most dominant GW signals, and bubble collision produce the least. These GW signals may lie  within the sensitivity reach of forthcoming detectors like the Big Bang Observer (BBO)~\cite{Harry:2006fi}, the DECi-hertz Interferometer Observatory (DECIGO)~\cite{Seto:2001qf}, Advanced LIGO (aLIGO)~\cite{Harry:2010zz}, and the Einstein Telescope (ET)~\cite{Punturo:2010zz}.
Following our objective, we consider an extension of the SM by a $U(1)_X$ gauge symmetry and a complex scalar singlet $\Phi$ that acquires a vacuum expectation value (VEV)~\cite{Oda:2015gna, Das:2016zue, Gola:2022nkg}, thereby spontaneously breaking the new symmetry and enabling the phase transition to proceed via a first-order phase transition (FOPT). In this setup, the nature of the FOPT - and hence the strength of the resulting stochastic gravitational wave (GW) signal - is primarily governed by the interplay among the portal coupling $\lambda_{H\Phi}$, the singlet self-coupling $\lambda_\Phi$, and the singlet mass parameter $\mu_\Phi$, which sets the scale of the singlet VEV. The portal interaction $\lambda_{H\Phi} |H|^2 |\Phi|^2$ in Eqn.~\eqref{eq:scalar-potential} plays a crucial role in shaping the finite-temperature effective potential, particularly by enhancing the cubic term that induces a potential barrier between the symmetric and broken phases. A suitably large $\lambda_{H\Phi}$ and an appropriately chosen singlet VEV can significantly increase the barrier height and the latent heat released during the transition, thereby amplifying the GW signal.
In addition, the $U(1)_X$ gauge coupling $g_{_X}$ affects the thermal mass corrections to the singlet field through its interaction with the $U(1)_X$ gauge boson, further influencing the dynamics of the phase transition. Thus, the parameters $\lambda_{H\Phi}$, $\lambda_\Phi$, the singlet VEV and $g_{_X}$ constitute the key handles for realizing a strong FOPT and generating a potentially observable GW signal from the early Universe. To ensure anomaly cancellation, we incorporate three generations of right-handed fermionic singlets, namely right-handed neutrinos (RHNs). Their presence enriches the phenomenology of the model by enabling neutrino mass generation via the type-I seesaw mechanism~\cite{Minkowski:1977sc, Mohapatra:1979ia, Yanagida:1979as, Gell-Mann:1979vob, Yanagida:1980xy}. Moreover, the $U(1)_X$ structure introduces new contributions to lepton mixing, potentially modifying the Dirac CP-violating phase and the Majorana phases. We compute the GW spectrum in the presence of the scalar singlet $\Phi$, the new gauge boson $Z_X$, and the RHNs. Additionally, the Yukawa interactions of RHNs with the lepton doublets and the Higgs can naturally accommodate baryogenesis via leptogenesis, providing a possible explanation for the observed matter-antimatter asymmetry in the Universe.\\

The remainder of this work is organized as follows. In Sec.~\ref{sec:Model-framework}, we introduce the theoretical framework of the model, outlining the anomaly cancellation conditions, the motivation to extend the content of SM particles, and presenting the full Lagrangian. In Sec.~\ref{sec:neutrino-sector}, we detail the neutrino sector, highlighting how Dirac and Majorana mass terms generate neutrino masses via the type-I seesaw mechanism. Sec.~\ref{sec:leptogenesis} presents the implications for thermal leptogenesis. The gravitational wave sector is presented in Sec.~\ref{sec:gw}, beginning with the discussion of the FOPT in Sec.~\ref{sec:FOPT} and the formulation of the gravitational wave sources in Sec.~\ref{sec:gw-spectrum}, followed by a sensitivity analysis and finally, we summarize and conclude in Sec.~\ref{sec:conclusion}.

\section{Model Framework}
\label{sec:Model-framework}
In this section, we describe the theoretical framework of the model. To address the mixing pattern in the lepton sector, we extend the SM gauge group, $SU(2)_L \times U(1)_Y$, by introducing an additional local $U(1)_X$ gauge symmetry. The charge of the SM fields under this new symmetry, denoted by $Q_X$, is defined as a linear combination of the hypercharge ($Y$) and the $B-L$ charge (baryon number minus lepton number) associated with each field:
\begin{align}
Q_X = x_Y \cdot Y + x_{B-L} \cdot (B - L).
\end{align}
However, the gauge group with $B-L$ extension is not anomaly free, and hence the $U(1)_X$ derived from it is also anomalous, unless $x_{B-L} = 0$. The following triangular anomalies must be canceled~\cite{Peskin:1995ev}:
\begin{center}
$[\mathrm{SU}(3)_c]^2 \times \mathrm{U}(1)_X \quad \quad \quad
[\mathrm{SU}(2)_L]^2 \times \mathrm{U}(1)_X \quad \quad \quad
[\mathrm{U}(1)_Y]^2 \times \mathrm{U}(1)_X$  \\[1.5ex]
$[\mathrm{U}(1)_X]^2 \times \mathrm{U}(1)_Y  \quad \quad \quad 
[\mathrm{U}(1)_X]^3 \quad \quad \quad
\text{Grav}^2 \times \mathrm{U}(1)_X$
\end{center}
In our model, the inclusion of three generations of RHN, $N_R^i$, ensures cancellation of this anomaly~\cite{Oda:2015gna, Das:2016zue}. The total anomaly contribution from all SM fermions and the three $N_R^i$ fields sums to zero, making the theory consistent with general covariance. The particle content is summarized in Table~\ref{tab:Particle-content}, wherein a complex scalar field $\Phi$, singlet under $SU(2)_L$, is introduced with hypercharge $Y = 0$ and $B-L$ charge $Q^\Phi_{B-L} = 2$, for the spontaneous breaking of the $U(1)_X$ symmetry,  through FOPT.  For simplicity, we take $x_Y = -\frac{32}{41}$ and $x_{B-L} = 1$~\cite{Oda:2015gna, Das:2016zue}, which allows us to neglect the mixing between the $U(1)_Y$ and $U(1)_{X}$ gauge symmetries. With this choice of $x_Y$, the RG evolution of the kinetic mixing parameter, $g_{_\mathrm{XY}}$  between the two gauge fields (defined in Eqn.~\ref{eq: Dmu-kinetic}) becomes proportional to the parameter itself. Consequently, setting the mixing parameter to zero at a given scale ensures that it remains zero at all energy scales, since its RG evolution also vanishes. This choice effectively renders $U(1)_Y$ and $U(1)_{X}$ orthogonal in the corresponding charge space.
\renewcommand{\arraystretch}{1.5}
\begin{table}[h!]
\begin{center}
\begin{tabular}{||c||c|c|c|c||}
\hline \hline
Field & $SU(2)_L$& $Y$ & $B - L$ & $Q_X$ \\
\hline \hline
$Q_L = (u_L, d_L)$ & $2$& $1/6$ & $1/3$ & $x_Y/6 + x_{B-L}/3$ \\
$u_R$ & $1$ &$2/3$ & $1/3$ & $2x_Y/3 + x_{B-L}/3$ \\
$d_R$ & $1$ & $-1/3$ & $1/3$ & $-x_Y/3 + x_{B-L}/3$ \\
$\ell_L = (\nu_L, e_L)$ &$2$ & $-1/2$ & $-1$ & $-x_Y/2 - x_{B-L}$ \\
$e_R$ & $1$ & $-1$ & $-1$ & $-x_Y - x_{B-L}$ \\
$N_R^i$ & $1$ & $0$ & $-1$ & $-x_{B-L}$ \\ \hline
$H$ & $2$ & $1/2$ & $0$ & $x_Y/2$ \\
$\Phi$ & $1$ & $0$ & $Q^{\Phi}_{B-L}$ & $x_{B-L} Q^{\Phi}_{B-L}$ \\
\hline\hline
\end{tabular}
\caption{Particle spectrum in the extended SM guage symmetry $SU(2)_L \times U(1)_Y \times U(1)_X$. }
\label{tab:Particle-content}
\end{center}
\end{table}
The Lagrangian contains the following terms including the familiar SM part:
\begin{align}
\mathcal{L} &= \mathcal{L}_{\text{SM}} + \mathcal{L}_{\text{kin}} + \mathcal{L}_{\text{Yukawa}} - V_{\rm tree}(H, \Phi).
\end{align}
The kinetic part of Lagrangian is given as,
\begin{align}
\mathcal{L}_{\text{kin}} &= |D_\mu H|^2 + |D_\mu \Phi|^2 + i \overline{N_R^i} \gamma^\mu D_\mu N_R^i - \frac{1}{4} F_Y^{\mu\nu} F_{Y\mu\nu} - \frac{1}{4} F_X^{\mu\nu} F_{X\mu\nu}- \frac{\kappa}{2} F_{Y}^{\mu\nu} F_{X\mu\nu},
\end{align}
where $H$ is the SM Higgs doublet, $F_Y^{\mu\nu}$ and $F^{\mu\nu}_X$ are the field strengths for the gauge groups $U(1)_Y$ and $U(1)_X$, respectively and, $\kappa$ defines the mixing strength. The requirement for positive kinetic energy implies that the kinetic coefficient satisfies $|\kappa| < 1$. Now, we perform field redefinition to obtain the orthogonal bases ($B_{Y}$,~$B_X$):
\begin{equation}
\begin{pmatrix}
{B}_Y \\
{B}_X
\end{pmatrix}
=
\begin{pmatrix}
0 & \kappa \\
1 & \sqrt{1-\kappa^2}
\end{pmatrix}
\begin{pmatrix}
F_Y \\
F_X
\end{pmatrix}.
\end{equation}
The covariant derivative $D_\mu$ includes the $\mathrm{U}(1)_X$ gauge interaction:
\begin{align}
D_\mu &\equiv \partial_\mu - i g \tau^a W_\mu^a - i g' {Y} B_{{_Y}\mu} - i \left(g_{_X} Q_X\frac{1}{\sqrt{1-\kappa^2}}-g_{_{XY}}Y\right) B_{_X\mu}
\label{eq: Dmu-kinetic}
\end{align}
where, $g_{_X}$ is the gauge coupling associated with $U(1)_{X}$ and $g_\mathrm{_{XY}}= g^\prime \frac{\kappa}{\sqrt{1-\kappa^2}}$. We take $g_{_{XY}} = 0$ to neglect the mixing in the gauge sector. Further, the Yukawa interaction is governed by the following terms:  
\begin{align}
\mathcal{L}_{\text{Yukawa}} \supset -\mathbf{Y^\ell_{\alpha\beta}}\overline{\ell_L^\alpha}He_R^\beta- \mathbf{Y^\nu_{\alpha i}} \overline{\ell_L^\alpha} \tilde{H} N_R^i - \frac{1}{2} \mathbf{Y^N_{ij}} \Phi \overline{(N_R^i)^c} N_R^j + \text{h.c.},
\label{eq:Yukawa}
\end{align}
where, $\tilde{H} = i \sigma_2 H^*$ and $y_{\alpha \beta}$, $y_{\alpha i}$, $y_{ij}$ denote the Yukawa couplings, with the indices running as $\alpha, \beta = e, \mu, \tau$ and $i, j = 1, 2, 3$. The scalar potential at tree level $V_{\mathrm{tree}}(H, \Phi)$ is discussed in detail as follows.
\section*{Scalar Sector}
The model includes a scalar singlet $\Phi$, which also contributes to the total tree level potential other than SM Higgs, as: 
\begin{align}
V_{\rm tree}(H, \Phi) = -\mu_H^2 |H|^2 + \lambda_H |H|^4 - \mu_\Phi^2 |\Phi|^2 + \lambda_\Phi |\Phi|^4 + \lambda_{H\Phi} |H|^2 |\Phi|^2,
\label{eq:scalar-potential}
\end{align}
where $\lambda_{H\Phi}$ indicates the coupling of Higgs $H$ with the $\Phi$. For our analysis, we set $\lambda_{H\Phi} = 0.1$.
The scalar fields acquire VEV:
\begin{align}
\langle H \rangle = \frac{1}{\sqrt{2}} \begin{pmatrix} 0 \\ v_{_H} \end{pmatrix}, \quad \langle \Phi \rangle = \frac{v_\Phi}{\sqrt{2}}.
\end{align}
The field $\Phi$ can be expressed as: 
\begin{align}
    \Phi = \frac{h_\Phi + v_\Phi}{\sqrt{2}} + i \frac{\chi}{\sqrt{2}}, 
\end{align}
where, $h_\Phi$ denotes the CP-even term and $\chi$ accounts for the CP-odd term. Taking into account the kinetic parts of the scalars as mentioned in Eqn.~\ref{eq: Dmu-kinetic}, we can write the mass of new gauge boson after the $U(1)_X$ symmetry breaking:
\begin{align}
    M_{Z_X}^2 = g_{_X}^2 Q_{\Phi}^2 v_\Phi^2.
    \label{eq: MZ'mass}
\end{align}

\section{Neutrino Sector}
\label{sec:neutrino-sector}
In the case of charged leptons, the Yukawa couplings \( y_{\alpha \beta} \) are taken to be real without any loss of generality, as any complex phase can be absorbed through suitable field redefinitions. This choice of Yukawa coupling yields physical masses of charged leptons. The charged lepton mass matrix, generated by the first term in Eqn. \ref{eq:Yukawa}, i.e., in $\mathcal{L}_{\text{Yukawa}}$ turns out as non-diagonal:
\begin{align}
    M_\ell = \begin{pmatrix}
        y_{ee}~~ &~~ y_{e\mu}~~ &~~y_{e\tau}\\
        y_{\mu e}&y_{\mu\mu}&y_{\mu\tau}\\
        y_{\tau e}&y_{\tau \mu}&y_{\tau\tau}
    \end{pmatrix} \frac{v_{_H}}{\sqrt{2}},
    \label{eq: Charged-lepton}
\end{align}
which can also be rewritten as,
\begin{align}
    M_\ell = \mathbf{Y^\ell_{\alpha\beta}}\frac{v_H}{\sqrt{2}}.\label{mm}
\end{align}
The charged lepton mass matrix  $M_{\ell}$ (\ref{eq: Charged-lepton})  can be diagonalized by introducing two independent unitary matrices, $V_L^\ell$ and $V_R^\ell$ such that, 
\[
V_L^{\ell \dagger} M_\ell V_R^\ell = \rm \textbf{Diag}\,(m_e, m_\mu, m_\tau)\, ,
\]
where, $m_e$, $m_\mu$, and $m_\tau$ are the physical masses of e, $\mu$ and $\tau$ respectively. 
For the determination of the model parameters, 
it is convenient to express the physical masses in terms of basis-independent quantities constructed from the mass matrix. More explicitly, the determinant of $M_\ell M_\ell^\dagger$ is given by the product of the squared charged-lepton masses,
\begin{equation}
\mathrm{Det}(M_\ell M_\ell^\dagger) = m_e^2 \, m_\mu^2 \, m_\tau^2,
\end{equation}
while the trace provides the sum of the squared masses,
\begin{equation}
\mathrm{Tr}(M_\ell M_\ell^\dagger) = m_e^2 + m_\mu^2 + m_\tau^2.
\end{equation}
Furthermore, the second invariant constructed from trace combinations,
\begin{equation}
\frac{1}{2} \left[ \left( \mathrm{Tr}(M_\ell M_\ell^\dagger) \right)^2 
- \mathrm{Tr}\left( (M_\ell M_\ell^\dagger)^2 \right) \right],
\end{equation}
corresponds to the sum of pairwise products of the squared masses,
\begin{equation}
m_e^2 m_\mu^2 + m_\mu^2 m_\tau^2 + m_\tau^2 m_e^2.
\end{equation}
These relations follow from the general properties of a $3 \times 3$ matrix.
Using these relations, the obtained 
values of $V_L^\ell$ and $V_R^\ell$ are presented in the appendix~\ref{app:Unitary matrices} for the set of Yukawa couplings as given in Table \ref{tab:Yukawa-values}.
\begin{table}[H]
\centering
\begin{tabular}{|c|c|c|c|c|c|c|c|c|}
\hline
$y_{ee}$ & $y_{e\mu}$ & $y_{e\tau}$ & $y_{\mu e}$ & $y_{\mu\mu}$ & $y_{\mu\tau}$ & $y_{\tau e}$ & $y_{\tau \mu}$ & $y_{\tau \tau}$ \\
\hline \hline 
6.87 & $6.58$ & $-1.28$ & $-2.34$ & $-1.45$ & $0.63$ & $-1.58$ & $-1.42$ & 0.32 \\
\hline
\end{tabular}
\caption{Yukawa couplings (rescaled as $y_{\alpha\beta}(\times 10^{-3})$)) in the charged lepton mass matrix.}
\label{tab:Yukawa-values}
\end{table}

Furthermore, we have the second term in the Lagrangian $\mathcal{L}_{\text{Yukawa}}$, which corresponds to a Dirac-type interaction and gives rise to the following mass matrix for the neutral part of lepton sector:
\begin{align}
    M_D = \begin{pmatrix}
        y_{e1}~~ &~~ y_{e2}~~ &~~y_{e3}\\
        y_{\mu 1}&y_{\mu2}&y_{\mu3}\\
        y_{\tau1}&y_{\tau2}&y_{\tau3}
    \end{pmatrix}\frac{v_{_H}}{\sqrt{2}}\equiv \mathbf{Y^\nu_{\alpha i}}\frac{v_H}{\sqrt{2}}.
\end{align}
Moreover, the Majorana mass matrix derived from last term of $\mathcal{L}_{\text{Yukawa}}$ is presented as follows:
\begin{align}
    M_N = \begin{pmatrix}
        y_{11}~~ &~~ y_{12}~~ &~~y_{13}\\
        y_{12}&y_{22}&y_{23}\\
        y_{13}&y_{23}&y_{33}
    \end{pmatrix}\frac{v_\Phi}{\sqrt{2}}\equiv \mathbf{Y^N_{ij}}\frac{v_\Phi}{\sqrt{2}}.
    \label{eq:MN}
\end{align}
Using type-I seesaw, we can write the light active neutrino mass matrix as:
\begin{align}
M_{\nu} = -M_D M_N^{-1}M_D^T.
\label{eq:numass}
\end{align}
We define,
\begin{align}
    \kappa = \frac{1}{v_H^2}\,M_\nu
\end{align}
    
This matrix can be diagonalized as
\begin{align}
V^{\nu T} \kappa V^\nu = \mathrm{diag}(\kappa_{1}, \kappa_{2}, \kappa_{3}) = D_\kappa
\end{align}
while the RHN mass matrix is diagonalized according to 
\begin{equation}
M_N = U_M^* D_N U_M^\dagger,
\qquad
D_N = \mathrm{diag}(M_{N_1}, M_{N_2}, M_{N_3}) \, .
\end{equation}

\begin{figure}
    \centering
    \includegraphics[width=0.5\linewidth]{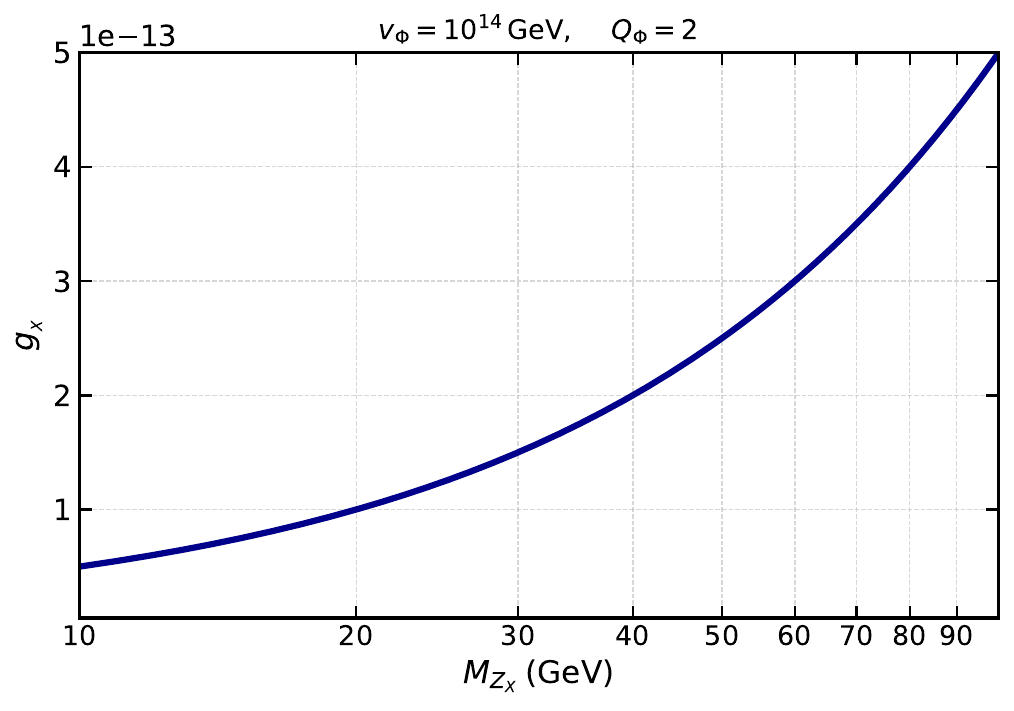}
    \caption{Mass of $Z_X$ vs the gauge coupling $g_{_X}$ for the VEV of $\Phi$ as $10^{14}$ GeV and $Q_\Phi$=2.}
    \label{fig:MZ-G'}
\end{figure}

Using the Casas-Ibarra parameterization~\cite{Casas:2001sr}, the neutrino Yukawa matrix can be written as
\begin{equation}
\mathbf{Y^\nu_{\alpha i}}
=
U_M^*\, D_{\sqrt{N}}\,
\mathbf{R}\,
D_{\sqrt{\kappa}}\,
V^{\nu\dagger},
\label{eq:CI}
\end{equation}
where $D_{\sqrt{x}} \equiv \sqrt{D_x}$ and $\mathbf{R}$ is a complex orthogonal matrix satisfying $\mathbf{R}^T\mathbf{R}=\mathbb{1}$.  
The unknown high-scale parameters are encoded in $\mathbf{R}$, and different choices of $\mathbf{R}$ correspond to different neutrino Yukawa textures. By scanning over $\mathbf{R}$, we identify benchmark Yukawa matrices $\mathbf{Y^\nu_{\alpha i}}$ that simultaneously reproduce the observed light neutrino masses and oscillation parameters from the Nufit-6.0~\cite{Esteban:2024eli}. We select five  benchmark sample points (SPs) that satisfy the current neutrino oscillation constraints. The associated Yukawa couplings $\mathbf{Y^\nu_{\alpha i}}$ and $\mathbf{R}$ matrices are presented in Table~\ref{tab:App-Ymatrix} and~\ref{tab:App-Rmatrix}, respectively. The Yukawa couplings associated with the Majorana mass term are given in Table~\ref{tab:MNyukawa}, which leads to the hierarchical structure of RHNs as $\left(M_{N_1}, M_{N_2}, M_{N_3}\right) =\left(10^{10},\,10^{11},\,10^{12}\right)\,\text{GeV}$. These benchmarks will serve as the basis for the subsequent analysis of leptogenesis and GW phenomenology. For the chosen VEV ($v_\Phi = 10^{14}$ GeV) we present the correlation of $M_{Z_X}$ vs $g_{_X}$ in Fig.~\ref{fig:MZ-G'}, using relation given in Eqn.~\ref{eq: MZ'mass}.

\section{Thermal Leptogenesis}
\label{sec:leptogenesis}

The RHNs not only explain the origin of neutrino masses but also play a key role in generating the observed baryon asymmetry of the Universe (BAU) via leptogenesis.  These RHNs are Majorana in nature and satisfy the condition $M_{N_i} > 10^{9}~\mathrm{GeV}$, to explain the BAU according to Ref.~\cite{Davidson:2002qv}. For successful leptogenesis, the three Sakharov conditions must be fulfilled~\cite{Sakharov:1967dj}: (i) lepton number violation through the decay of RHNs ($N_i \rightarrow \ell H$), (ii) $C$ and $CP$ violating Yukawa interactions of RHNs, and (iii) departure from thermal equilibrium during RHN decay. The CP asymmetry is mainly coming from,
\begin{align}
\varepsilon_i
&=
\frac{
\Gamma(N_i \to \ell H)
-
\Gamma(N_i \to \bar{\ell} H^\ast)
}{
\Gamma(N_i \to \ell H)
+
\Gamma(N_i \to \bar{\ell} H^\ast)
} \, .
\end{align}
For decay of lightest RHN, $N_1$, the CP asymmetry parameter is given as~\cite{Okada:2025daq},
\begin{align}
\varepsilon_1= -\frac{3M_{N_1}}{8\pi v_H^2}\, \frac{\sum_{j = 2, 3}
\operatorname{Im} [( \mathbf{R} {D}_\nu \mathbf{R}^\dagger)_{1j}^2]}{( \mathbf{R} {D}_\nu \mathbf{R}^\dagger)_{11}},
\end{align}
The heavy neutrino and lepton asymmetry evolution is defined as~\cite{Plumacher:1996kc},
\begin{align}
\frac{dY_1}{dx} &= 
- \frac{x}{s(M_{N_1}) H(M_{N_1})}
\left( \frac{Y_1}{Y_1^{\mathrm{eq}}} - 1 \right)
\left( \gamma_D + \gamma_S \right),
\\[1em]
\frac{dY_L}{dx} &= 
\frac{x}{s(M_{N_1}) H(M_{N_1})}
\left[
\left( \frac{Y_1}{Y_1^{\mathrm{eq}}} - 1 \right)
\epsilon_1 \gamma_D
-
\left( \frac{Y_L}{Y_\ell^{\mathrm{eq}}} \right)
\gamma_W
\right]\, ,
\end{align}
with the Hubble expansion rate $H(M_{N_1})$ and entropy density $s(M_{N_1})$,
\begin{align}
H(M_{N_1}) &=
\left( \frac{\pi^2 g_*^{\mathrm{}}}{90} \right)^{1/2}
\frac{M_{N_1}^2}{M_P}\, ,
\\[1em]
s(M_{N_1}) &= 
\frac{2\pi^2}{45}
g_*^{\mathrm{SM}}
M_{N_1}^3\, .
\end{align}
Here, $Y_1$ denotes the yield of the RHN $N_1$, 
while $Y_1^{\mathrm{eq}}$ represents its equilibrium yield. 
The quantity $Y_L$ corresponds to the lepton asymmetry yield, 
and $Y_L^{\mathrm{eq}}$ is the equilibrium yield of the lepton doublets. The yield is defined as $Y_A = n_A/s$, where $n_A$ is the number density of species $A$ and $s$ is the entropy density. The variable $x = M_{N_1}/T$ is a dimensionless time parameter. Furthermore, $\gamma_D$ denotes the decay reaction density, $\gamma_S$ represents the $\Delta L = 1$ scattering reaction density, and $\gamma_W$ corresponds to the total washout reaction density. The quantity $g_*^{\mathrm{SM}}$ is the effective number of relativistic degrees of freedom evaluated at $T = M_{N_1}$. The generated lepton asymmetry can be subsequently transferred into baryon asymmetry via sphaleron processe as,
\[
Y_B = - C_{\mathrm{sph}} \, Y_L\, ,
\]
with $C_{\mathrm{sph}} = 28/79$. 
Using the relevant Yukawa couplings obtained from  section~\ref{sec:neutrino-sector} and presented in Table~\ref{tab:App-Ymatrix}, we numerically solve the Boltzmann equations governing the evolution of the RHN abundance and the lepton asymmetry. We find that the same benchmark Yukawa couplings that successfully reproduce the observed neutrino masses and mixing angles can also account for the observed BAU. For each benchmark point the values obtained for $\varepsilon_!$ and $Y_B$, presented in Table~\ref{tab:lepto}, lie within the required range to account for the observed baryon asymmetry of the Universe.

\begin{table}[h]
\centering
\renewcommand{\arraystretch}{1.2}
\begin{tabular}{|c|c|c|c|c|c|}
\hline
 & \textbf{SP1} & \textbf{SP2} & \textbf{SP3} & \textbf{SP4} & \textbf{SP5} \\
\hline
${\varepsilon_1 \; (\times 10^{-8})}$ 
& -5.6226 
& -5.6408 
& -5.6550 
& -5.6999 
& -5.6952 \\
\hline
${Y_B \; (\times 10^{-11})}$ 
& 8.6241 
& 8.6521 
& 8.6738 
& 8.7427 
& 8.7355 \\
\hline
\end{tabular}
\caption{CP asymmetry parameter $\epsilon_1$ and baryon asymmetry $Y_B$ for selected sample points.}
\label{tab:lepto}
\end{table}

\section{Gravitational Waves}
\label{sec:gw}
This section outlines the theoretical framework necessary to realize a FOPT in a $U(1)$ extension. The analysis incorporates one-loop corrections  to the tree-level potential at both zero and finite temperature. All these contributions combine to form the total effective potential. We initiate the GW analysis by introducing the relevant formulations, subsequently discussing contributions to the GW spectrum from different sources during FOPT, and conclude with a sensitivity analysis of obtained GW spectrum. (add in appendix:)

\subsection{Phase Transition Dynamics}
\label{sec:FOPT}
Here, we present how the GW generation can be studied in an additional $U(1)_X$ framework. The tree-level SM Higgs potential alone is not sufficient to realize a FOPT at the electroweak scale with a 125~GeV Higgs boson. To address this, we introduce a scalar field $\Phi$ associated with the $U(1)_X$ symmetry. However, achieving the desired FOPT still requires incorporating one-loop corrections to the tree-level potential. The one-loop quantum correction to $V_{\text{tree}}(H, \Phi)$ at zero temperature is given by the Coleman-Weinberg (CW) potential~\cite{Coleman:1973jx}. In our analysis, we work in the Landau gauge and adopt the $\overline{\text{MS}}$ renormalization scheme~\cite{Peskin:1995ev}:
\begin{align}
V_{\text{CW}}(\phi) = \sum_i \frac{n_i}{64\pi^2} m_i^4(\phi) \left( \log\frac{m_i^2(\phi)}{\mu^2} - c_i \right),
\end{align}
where $n_i$ is the degree of freedom, $m_i(\phi)$ is the field-dependent mass, $\mu$ is the renormalization scale, and $c_i = 3/2$ (scalars, fermions) or $5/6$ (gauge bosons). Further, to move towards the EW scale, we need to add finite temperature part of the one loop corrections to this potential:
\begin{align}
V_T(\phi, T) = \sum_i \frac{n_i T^4}{2\pi^2} J_{B/F} \left(\frac{m_i^2(\phi)}{T^2}\right),
\end{align}
where $J_{B/F}$ are thermal integrals for bosons/fermions~\cite{Kapusta:2006pm, Bellac:2011kqa}:
\begin{align}
J_B(x^2) &= \int_0^\infty dy\, y^2 \log\left[1 - e^{-\sqrt{y^2 + x^2}}\right], \\
J_F(x^2) &= \int_0^\infty dy\, y^2 \log\left[1 + e^{-\sqrt{y^2 + x^2}}\right].
\end{align}
The infrared divergence in one-loop finite-temperature corrections, arising from nearly massless bosonic modes, can be regulated through daisy resummation~\cite{Delaunay:2007wb, Espinosa:2007qk, Profumo:2007wc, Noble:2007kk, Espinosa:2008kw, Espinosa:2011ax, Curtin:2014jma, Blinov:2015sna, Basler:2016obg, Basler:2017uxn, Chala:2018ari}:
\begin{align}
V_{\text{daisy}}(\phi, T) = -\frac{T}{12\pi} \sum_{i\rm{=}bosons} n_i \left[ \left( m_i^2(\phi) + \Pi_i(T) \right)^{3/2} - \left( m_i^2(\phi)\right)^{3/2} \right],
\end{align}
where $\Pi_i(T)$ is the thermal (Debye) mass correction for $i$th boson mode. The Debye correction is zero for the transverse mode of gauge of gauge bosons. For longitudinal mode of gauge boson, Debye correction is given as:
\begin{align}
\Pi_{Z_X}(T) = \frac{1}{3} g_{_X}^2  T^2, \quad \quad
\Pi_{\rm scalar}(T) = \left( \frac{1}{4}g_{_X}^2 + \frac{1}{6}\lambda_\Phi \right) T^2\, .
\end{align}
For a thorough calculation of the field-dependent masses, please refer to Appendix \ref{app:fielddependentmass}. Since one-loop corrections are added to the tree-level potential, a counter-term potential ($V_{CT}(\phi)$) must be included to absorb the divergences and preserve the renormalizability of the theory, thereby keeping physical observables intact. The detailed derivation of the counter-term potential is given in Appendix \ref{app:counterterms}. The complete finite-temperature effective potential is:
\begin{align}
V_{\text{eff}}(\phi, T) = V_0(\phi) + V_{\text{CW}}(\phi) + V_T(\phi, T) + V_{\text{daisy}}(\Phi, T)+ V_{\rm CT}(\phi).
\label{eq:Veff}
\end{align}
The discussion begins with a model-independent overview and then proceeds to the specific potential considered in the Eqn~\ref{eq:Veff}. The stochastic GW background is characterized by:

\begin{itemize}
\item \textbf{Transition temperature ($T_\star$):} As the Universe cools down to electroweak epoch, $V_{\mathrm{eff}}$ starts building a new minimum at non-zero value of $\phi = \phi_{\mathrm{low}}$ (broken phase), under a FOPT. The corresponding transition or critical temperature can be obtain from the following condition:
\begin{align} 
    V_{\mathrm{eff}}(\phi,~T)|_{(\phi_{\mathrm{high}},~T_\star)} = V_{\mathrm{eff}}(\phi,~T)|_{(\phi_{\mathrm{low}},~T_\star)},
\end{align}
where, $\phi_\mathrm{{high}}$ corresponds to the symmetric phase.

\item \textbf{Nucleation temperature ($T_n$):} Bubble nucleation remains suppressed at $T = T_\star$, but at $T_n~(>T_\star)$ bubble formation and subsequent expansion starts~\cite{Huber:2008hg}. Mathematically, the bubble nucleation rate per unit volume per unit time $ \Gamma|
_{T_n} \sim H|_{T_n}$, where $H$ is Hubble parameter.

\item \textbf{Percolation temperature $T_\mathrm{p}$:} At this temperature, Universe become dominated by the true vacuum. Percolation temperature is required if we are interested in supercooled phase transition. This leads to a new reheating phase, which is cosmologically very interesting to explore.

\item \textbf{The latent heat $\alpha$:}
    \begin{align}
        \alpha =  \frac{1}{\rho_{\mathrm{rad}}}\left[ \left( V - T \left. \frac{\partial V}{\partial T} \right|\right )_{\{ \phi_{\text{high}},\, T_n\}} - \left( V - T \left. \frac{\partial V}{\partial T} \right|\right )_{\{ \phi_{\text{low}},\, T_n\}}\right ],
    \end{align}
where, the radiation energy density:
\[
\rho_{\text{rad}} = \frac{\pi^2}{30} g_* T^4,
\]
with $g_*$ as the total number of relativistic degrees of freedom in the thermal plasma.
\item \textbf{The inverse duration $\beta/H_\star$:} When the temperature drops below $T_\star$, the new minimum of the potential shifts toward a nonzero value of $\phi$. This triggers a first-order phase transition, during which the Universe evolves from the symmetric phase to the broken phase via the nucleation of bubbles.  The bubble nucleation probability per unit volume and per unit time is equivalent to the decay rate of the false vacuum ($\Gamma(T)$). We can define $\Gamma(T)$ in two different conventions, which depend on temperature regime and method of calculations. In general, we can write,
\begin{align}
\Gamma(T) = \Gamma_0(T)\, e^{-S(T)}
\end{align}
and specifically, in high temperature regime,
\begin{align}
\Gamma(T) = \Gamma_0(T)\, e^{-{S_3(T)}/{T}}
\end{align}
where, $\Gamma_0(T)$ is the temperature dependent prefactor, $S(T)$ is the four-dimensional Minkowski action and  $S_3(T)$ is the three dimensional Ecucledian action~\cite{Turner:1992tz}. Now, the inverse of time scale of phase transition is:
\[
\frac{\beta}{H_\star} \simeq T \left. \frac{dS(T)}{dT}  \right|_{T = T_n} = T \left. \frac{d}{dT} \left( \frac{S_3(T)}{T} \right) \right|_{T = T_n}.
\]

\end{itemize}
\begin{figure}[h!]
    \centering
    \includegraphics[width=0.82\textwidth]{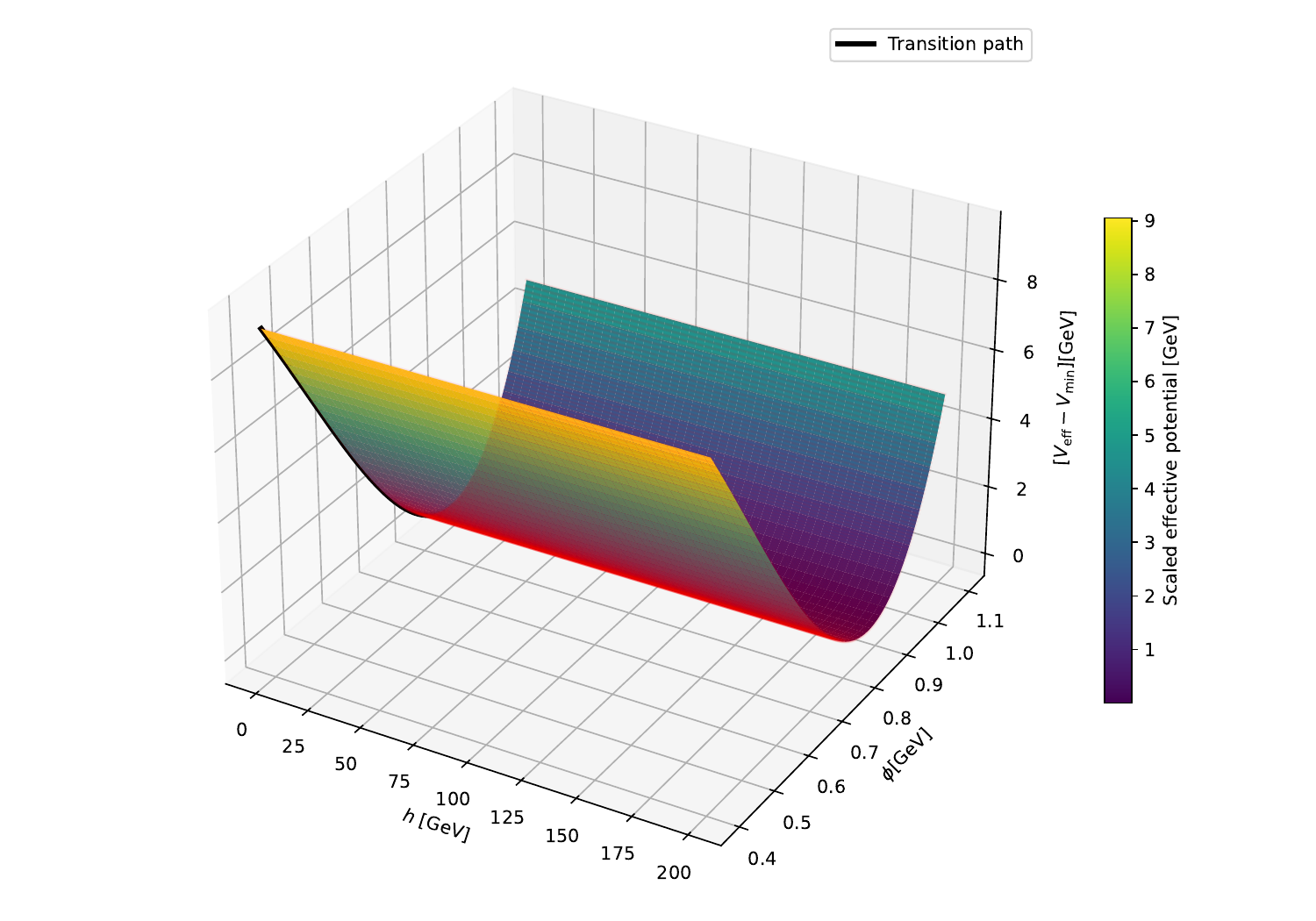}
    \caption{
Finite-temperature effective potential $V_{\rm eff}(h,\phi,T)$ evaluated at the nucleation temperature $T_n \simeq 2~\mathrm{TeV}$ as a function of the Higgs field $h$ and the singlet field $\phi$, shifted by its global minimum and
normalized by $T_n^4$.  }
    \label{fig:3dplot}
\end{figure}
Figure~\ref{fig:3dplot} displays the three-dimensional finite-temperature effective potential 
$V_{\rm eff}(h,\phi,T)$ evaluated at the nucleation temperature 
$T_n \simeq 2~\mathrm{TeV}$ as a function of the Higgs field $h$ and the singlet field $\phi$. 
The potential is shifted by its global minimum and normalized by $T_n^4$ 
to emphasize its structure in the phase-transition region. 
The surface exhibits a nontrivial minimum at nonzero $\phi$ separated from the symmetric region by a potential barrier. 
The presence of this barrier structure at $T_n$ reflects the first-order nature of the phase transition.

\subsection{Gravitational Wave Spectrum}
\label{sec:gw-spectrum}
The GW power spectrum from three main mechanisms: bubble collisions, sound waves, and turbulence is,
\begin{align}
h^2\Omega_{\text{GW}}(f) = h^2\Omega_{\text{col}} + h^2\Omega_{\text{sw}} + h^2\Omega_{\text{turb}},
\end{align}
where each term depends on $\alpha$, $\beta/H_\star$, $T_n$ and the bubble wall velocity $v_\mathrm{w}$. Moreover, these mechanisms are discussed in further detail below.
\subsubsection{Bubble collision}
Among the three mechanisms discussed above, the gravitational waves produced from bubble wall collisions are the weakest in amplitude. The corresponding power spectrum is given by~\cite{Huber:2008hg}:
\begin{align}
h^2\Omega_\mathrm{col}(f)  &= h^2\Omega_\mathrm{col}(f^\mathrm{col}_\mathrm{p}) \, \mathcal{S}_\mathrm{col}(f), 
\end{align}
where, the spectral peak component~\cite{Kosowsky:1992vn},
\begin{align}
h^2 \Omega_{\text{col}}^{}(f^{\mathrm{col}}_{\text{p}}) \simeq 1.7 \times 10^{-5} \, \kappa^2_{\text{col}} \, \Delta \left( \frac{\beta}{H_\star} \right)^{-2} \left( \frac{\alpha}{1 + \alpha} \right)^2 \left( \frac{100}{g_\ast} \right)^{1/3}
\end{align}
with peak frequency
\begin{align}
f^{\mathrm{col}}_{\text{p}} \simeq   \left( \frac{10.54 \times 10^{-6}}{1.8 - 0.1\, v_\mathrm{w}+ v_\mathrm{w}^2} \right) \left( \frac{\beta}{H_\star} \right) \left( \frac{T_n}{100\, \text{GeV}} \right) \left( \frac{g_\ast}{100} \right)^{1/6} \, \text{Hz}
\end{align}
The spectral profile is defined as \cite{Huber:2008hg}:
\[
\mathcal{S}_{\text{col}}(f) =\frac{ (x+y)\,(f_\mathrm{p}^{\mathrm{col}})^y\,f^x}{y\,(f_\mathrm{p}^{\mathrm{col}})^{x+y}\,+\,x\,f^{x+y}}
\]
with $x = 2.7$ and $y = 1.0$. The fitting function describing the spectrum variation with the bubble velocity~\cite{Kamionkowski:1993fg},
\[
\Delta = \frac{0.11 v_\mathrm{w}^3}{0.42+v_\mathrm{w}^2},
\]
and efficiency $\kappa_{\mathrm{col}}$~\cite{Kamionkowski:1993fg}:
\[
\kappa_{\text{col}} = \frac{1}{1+0.715\alpha}\left(0.715\alpha +\frac{4}{27}\sqrt{\frac{3\alpha}{2}}   \right)
\]
\subsubsection{Sound waves}
The pressure generated from the energy difference between true and test vacua per unit volume, propagates in the form of GW, with a comparatively strong strength. The mathematical form of such spectrum is given as:
\begin{align}
h^2\Omega_\mathrm{sw}(f)  &= h^2\Omega_\mathrm{sw}(f^\mathrm{sw}_\mathrm{p}) \, \mathcal{S}_\mathrm{sw}(f), 
\end{align}
with spectral peak~\cite{Hindmarsh:2013xza,Hindmarsh:2015qta,Espinosa:2010hh}
\begin{align}
h^2\Omega_\mathrm{sw}(f^\mathrm{sw}_\mathrm{p} ) &\simeq 2.65 \times 10^{-6} \, v_\mathrm{w} \,\Upsilon \, \left (\frac{\beta}{H_\star}\right )^{-1} 
\left( \frac{ \alpha\,\kappa_{\text{s}}}{1 + \alpha} \right)^2 
\left( \frac{g}{100} \right)^{1/3}
\label{eq:sw_amp}
\end{align}
where, peak frequency~\cite{Huber:2008hg}
\begin{align}
f^\mathrm{sw}_\mathrm{p} &\simeq 1.9 \times 10^{-5} \left(\frac{1}{v_\mathrm{w}} \right) \left(\frac{\beta}{H_\star}\right ) 
\left( \frac{T_n}{100} \right) 
\left( \frac{g_*}{100} \right)^{1/6} \mathrm{Hz} ,
\label{eq:sw_peak} 
\end{align}
\begin{align}
    \Upsilon = 1 - \frac{1}{\sqrt{1 + 2\,\tau_{\mathrm{sw}} H_s}}
\end{align}
The quantity $H_s = 0.67$, refers to the Hubble rate corresponding to the onset of the source activity.  The time interval over which sound waves are actively produced is given by~\cite{Hindmarsh:2020hop, Guo:2020grp}:
\begin{align}
    \tau_{\mathrm{sw}} =\frac{ R_\star}{U_f}, \quad\quad U_f = \sqrt{\frac{3\,\alpha\,\kappa_{\mathrm{s}}}{4}}, \quad \mathrm{and} \quad R_\star = \frac{(8\pi)^{1/3} v_{\rm w} H_\star}{\beta}
\end{align}
where, $R_{\star}$ is the mean bubble separation , $U_f$ is the root mean square fluid velocity. The efficiency factor for $v_\mathrm{w} \simeq 1$~\cite{Ellis:2020nnr, Espinosa:2010hh},
\begin{align}
\kappa_{\mathrm{s}} = \frac{\alpha}{0.73 + 0.083\sqrt{\alpha} + \alpha}
\end{align}
and the spectral function~\cite{Caprini:2015zlo} 
\begin{align}
\mathcal{S}_{\mathrm{sw}}(f) = \left( {f^\mathrm{sw}_\mathrm{p}}/{f} \right)^4 \left( \frac{7}{4 \left( {f^\mathrm{sw}_\mathrm{p}}/{f} \right)^2 + 3} \right)^{7/2}
\end{align}

\subsubsection{Turbulence}
When the Reynolds number is large enough, it implies the dominance of inertial effects over viscous damping, thereby inducing turbulence in the plasma, which is a key mechanism for producing GW with notable amplitude. The spectrum is described by
\begin{align}
h^2\Omega_\mathrm{turb}(f)  &= h^2\Omega_\mathrm{turb}(f^\mathrm{turb}_\mathrm{p}) \, \mathcal{S}_\mathrm{turb}(f), 
\end{align}
with peak amplitude~\cite{Kamionkowski:1993fg},
\begin{align}
h^2\Omega_{\mathrm{turb}}(f^\mathrm{turb}_\mathrm{p}) = 
3.35 \times 10^{-4}\, v_\mathrm{w} \, \left(\frac{\beta}{H_\star}\right ) ^{-1}  \left( \frac{\alpha \, \kappa_{\text{turb}}}{1 + \alpha} \right)^{3/2} \left( \frac{100}{g} \right)^{1/3},
\end{align}
and the corresponding peak frequency
\begin{align}
f^\mathrm{turb}_{\mathrm{p}}= 2.7 \times 10^{-5} \frac{1}{v_\mathrm{w}}  \left(\frac{\beta}{H_\star}\right )\left( \frac{T_n}{100~{\rm GeV}} \right) \left( \frac{g_*}{100} \right)^{1/6}.
\end{align}
We can write the spectral function in this scenario as~\cite{Caprini:2009yp, Caprini:2015zlo, Binetruy:2012ze},
\begin{align}
\mathcal{S}_\mathrm{turb}(f) &= \frac{\left({f}/{f^\mathrm{turb}_{\mathrm{p}}} \right)^3}{
 (1 + {f}/{f^\mathrm{turb}_\mathrm{p}} )^{11/3}(1+\frac{8\pi f}{h_{*}})  } ,
 \end{align}
 with
 
 \[
h_{*} = 16.5 \left(\frac{T_{n}}{100~\mathrm{GeV}}\right)\left(\frac{g_*}{100}\right)^{1/6} \mathrm{Hz}
\]
The efficiency factor associated with turbulence is taken to be one-tenth of the efficiency of the bubble collision contribution.

\subsection{GW Sensitivity Analysis}
\label{sec:gw-sens}

We now compare the predicted stochastic gravitational wave (GW) signals of the model with the projected sensitivities of current and future interferometers, including LISA~\cite{amaro2017laser}, BBO~\cite{Harry:2006fi}, DECIGO~\cite{Kawamura:2011zz}, ET, CE, and HLVK~\cite{Schmitz:2020syl}. The phase transition parameters are computed from the full finite-temperature effective potential using the \texttt{CosmoTransitions}~\cite{Wainwright:2011kj} package. For each benchmark point (BP), we determine the critical temperature $T_c$, the nucleation temperature $T_n$, the strength parameter $\alpha$, and the inverse duration parameter $\beta/H_\star$. These quantities fully characterize the amplitude and peak frequency of the GW spectrum. Throughout the analysis, we assume a relativistic bubble wall velocity $v_w \sim 0.9$. The scalar sector parameters and phase transition parameters $\alpha$, $\beta/H_\star$, and $T_n$ are listed in table \ref{tab:benchmark} and $g_X = 0.15$, all of which remain within perturbative and tree-level unitarity limits. The five benchmark points are characterized by the different set of $\alpha$, $\beta/H_\star$, $\lambda_H$, $\lambda_\Phi$ and $T_n$ values.

\begin{figure}[h!]
    \centering
    \includegraphics[width=0.72\textwidth]{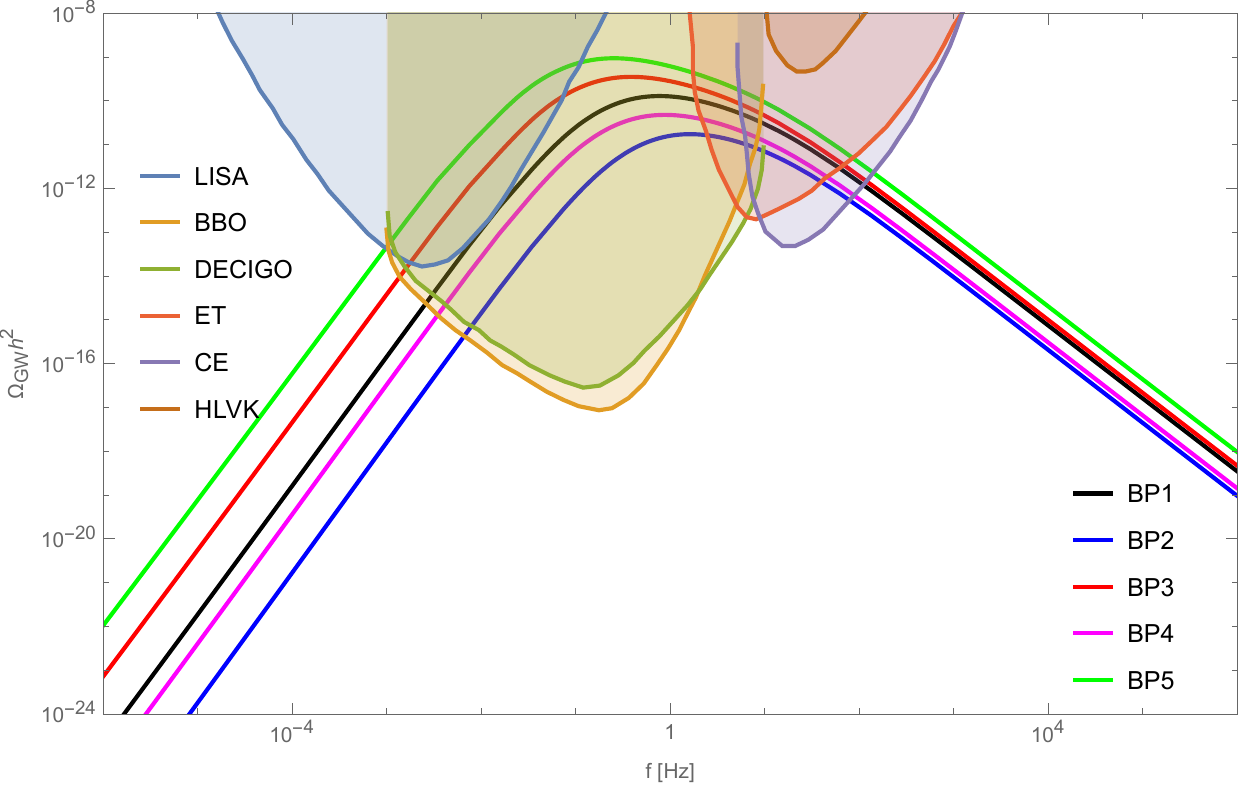}
    \caption{Predicted stochastic gravitational wave spectra for the five benchmark points compared with the projected sensitivity curves of LISA, BBO, DECIGO, ET, CE, and HLVK. The shaded regions indicate the expected experimental sensitivity.}
    \label{fig:GW_Sensitivity}
\end{figure}

The corresponding GW spectra are shown in Fig.~\ref{fig:GW_Sensitivity}, together with the projected sensitivity curves of the various experiments. The total GW signal includes contributions from bubble wall collisions, sound waves, and magnetohydrodynamic turbulence, with the sound-wave component typically providing the dominant contribution in the parameter region considered.

\begin{table}[h!]
\centering
\begin{tabular}{|c|c|c|c|c|c|c|}
\hline
\textbf{BPs} 
& $\boldsymbol{\alpha}$ 
& $\boldsymbol{\beta/H_\star}$ 
& $\boldsymbol{T_n~[\mathrm{GeV}]}$ 
& $\boldsymbol{\lambda_H}$
& $\boldsymbol{\lambda_\Phi}$
& $\boldsymbol{\lambda_{H\Phi}}$
\\
\hline \hline
BP1 & 0.567 & 413.60 & 2144.20 & 0.128 & 0.0020 & 0.10 \\
BP2 & 0.325 & 525.61 & 2588.00 & 0.130 & 0.0030 & 0.10 \\
BP3 & 0.624 & 236.10 & 1877.40 & 0.127 & 0.0015 & 0.10 \\
BP4 & 0.435 & 470.18 & 2272.62 & 0.131 & 0.0025 & 0.10 \\
BP5 & 0.825 & 148.18 & 1852.03 & 0.126 & 0.0008 & 0.10 \\
\hline
\end{tabular}
\caption{Phase transition parameters and scalar quartic couplings corresponding to the five benchmark points. The portal coupling is fixed at $\lambda_{H\Phi}=0.1$ and $v_\phi = 10^{14}$ GeV. All quartic couplings remain within perturbative and tree-level unitarity bounds.}
\label{tab:benchmark}
\end{table}

From Fig.~\ref{fig:GW_Sensitivity}, we observe that the predicted GW signals for all benchmark points fall within the projected sensitivity reach of DECIGO and BBO. Several benchmark points are also accessible to ET and CE, while none are detectable by HLVK due to its limited sensitivity in the relevant frequency band. The partial overlap with the LISA sensitivity curve reflects the fact that a TeV-scale phase transition shifts the peak frequency toward higher values, favoring deci-Hz detectors. These results demonstrate that the TeV-scale symmetry-breaking dynamics responsible for neutrino mass generation in this framework can produce observable stochastic gravitational wave signals, rendering the model testable in upcoming gravitational wave experiments.

We recompute the expected signal-to-noise ratios (SNR) for the benchmark points across a range of planned detectors, including LISA, DECIGO, BBO, ET, CE, and the HLVK network. The SNR is evaluated as
\begin{equation}
\text{SNR} = \sqrt{ \mathcal{T} \int_{f_{\rm min}}^{f_{\rm max}} df
\left( \frac{\Omega_{\rm GW}(f)}{\Omega_{\rm noise}(f)} \right)^2 },
\end{equation}
where $\mathcal{T}$ is the observation time for each detectors with the accessible frequency range  [$f_{\rm min}$, $f_{\rm max}$]. For the space-based interferometers, we include astrophysical foregrounds such as unresolved white dwarf binaries and extragalactic backgrounds in the effective noise, following \cite{Schmitz:2020syl}. 

\begin{figure}[h!]
    \centering
    \includegraphics[width=1.1\textwidth]{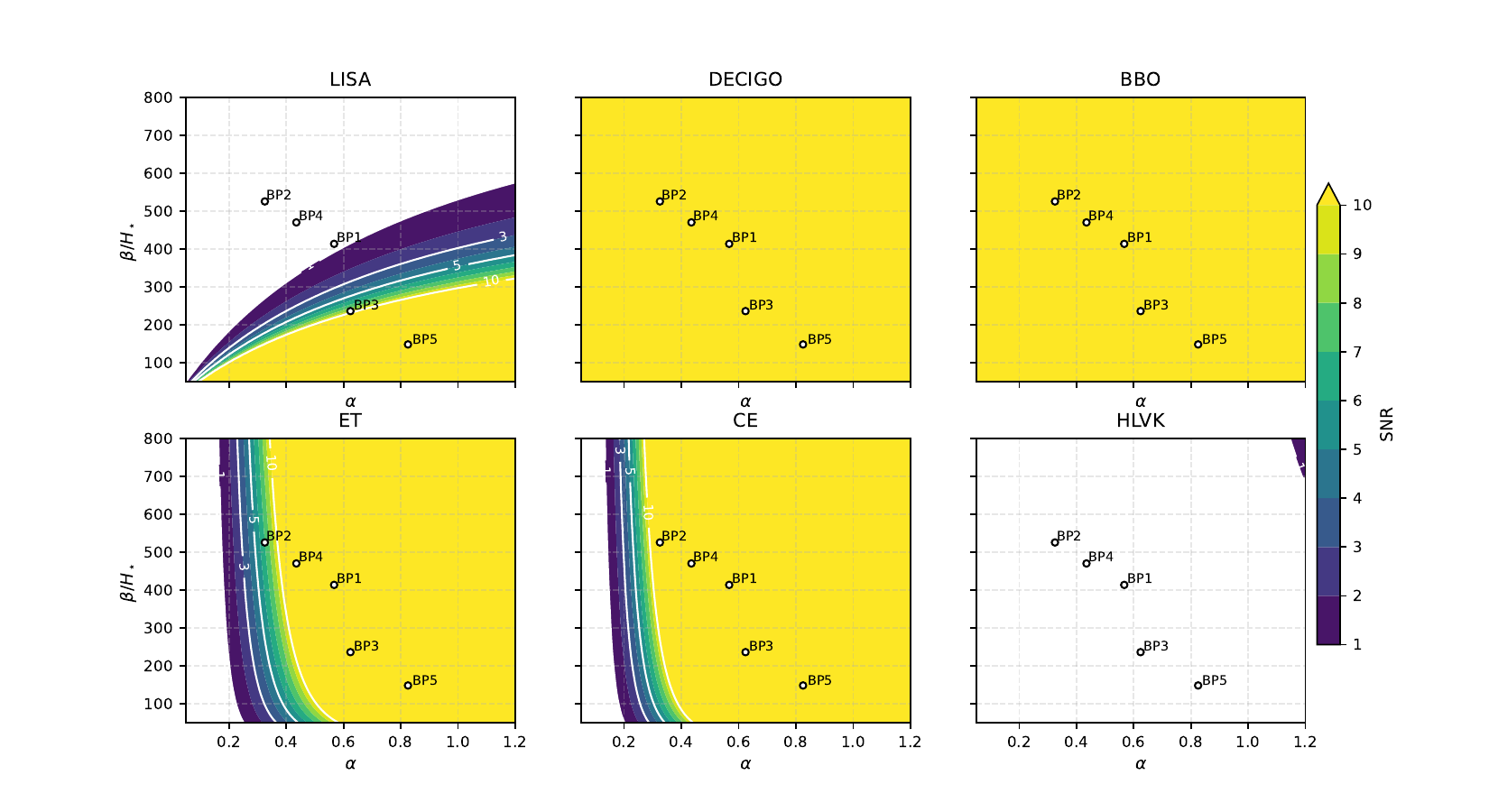}
    \caption{Signal-to-noise ratios (SNR) for the five benchmark points (BP1–BP5) across space-based interferometers and ground-based detectors. For LISA, astrophysical foregrounds from unresolved binaries are included in the effective sensitivity. The plots show that BP3 and BP5 yield large SNR values in space-based detectors, reaching $\text{SNR} \gtrsim 10$.}
    \label{fig:GW_SNR}
\end{figure}

Our updated analysis corresponds to the benchmark points given in tabel~\ref{tab:benchmark} with stronger phase transitions and $v_\mathrm {w} \sim 0.9$. The SNR contour plots are presented in Fig.~\ref{fig:GW_SNR}, where we present the SNR data for next generation of space-based interferometers: LISA, DECIGO and BBO in the top row and the ground-based detectors: ET, CE and HLVK in the bottom row. The SNR values are represented by the color scale shown on the right, with a detection threshold set at $\gtrsim 10$. Signals exceeding this threshold are considered to lie within the sensitivity range of the respective experiments. It is evident from the figure that, for LISA, the GW signals corresponding to BP1, BP2, and BP4 lie below the detection threshold, with BP3 sitting just at the limit, while BP5 appears within reach. In contrast, for DECIGO, BBO, and CE, all five benchmark points exceed the detection threshold.  For the ET, all benchmark points lie above the detection threshold except BP2, whereas for the HLVK, none of the benchmark points reach the detection limit. Notably, BP3 and BP5 generate signals accessible to future space-based interferometers, making them strong candidates consistent with collider and electroweak bounds.

\section{Conclusion}
\label{sec:conclusion}

In this work, we have presented a minimal and well-motivated extension of the Standard Model featuring a local $U(1)_X$ gauge symmetry and a scalar singlet field, whose spontaneous symmetry breaking triggers a strong first-order phase transition. The inclusion of three RHNs ensures anomaly cancellation and naturally realizes light active neutrino masses through a type-I seesaw mechanism. The Yukawa coupling matrix is reconstructed using the Casas-Ibarra parameterization, which inherently satisfies the current neutrino oscillation data. Additionally, the heavy right-handed neutrinos facilitate successful thermal leptogenesis, offering a consistent explanation for the observed baryon asymmetry.\\
We have demonstrated that the portal coupling and gauge interactions can significantly strengthen the phase transition, leading to a stochastic gravitational wave background detectable by upcoming experiments such as LISA, DECIGO, BBO, and the Einstein Telescope. We also present the SNR values for each benchmark point across the GW observatories. Our results show that all benchmark points are within the detection reach of DECIGO, BBO, ET, and CE, while none can be probed by HLVK. At LISA, only BP3 and BP5 fall above the detection threshold. \\
Altogether, our results establish a coherent framework linking neutrino phenomenology, baryogenesis, and gravitational wave cosmology. This scenario provides clear experimental targets for both cosmological and laboratory searches, highlighting the complementarity between future gravitational wave observatories and neutrino experiments in probing new physics beyond the Standard Model.

\section{Acknowledgement}
The work of AC was supported by the Japan Society for the Promotion of Science (JSPS) as a part of the JSPS Postdoctoral Program (Standard), grant number JP23KF0289. PM wants to thank Prime Minister's Research Fellows (PMRF) scheme for its financial support.
\appendix
\section{Unitary Matrix}
\label{app:Unitary matrices}
The unitary matrix $V_L^\ell$ required to diagonalize the charged lepton mass matrix $M_\ell$:
\[
V^\ell_L = 
\begin{pmatrix}
-0.939361 & 0.270655 & 0.210588 \\
-0.288029 & -0.955974 & -0.056148 \\
\;\;0.186120 & -0.113399 & 0.975961
\end{pmatrix}
\]
and 
\[
V^\ell_R =
\begin{pmatrix}
-0.726035 & -0.673045 & \;\;0.141011 \\
\;\;0.571402 & -0.704559 & -0.420828 \\
\;\;0.382586 & -0.224962 & \;\;0.896114
\end{pmatrix}
\]


\begin{table}[h]
\centering
\renewcommand{\arraystretch}{1.3}
\setlength{\arraycolsep}{8pt}
\begin{tabular}{|c|c|}
\hline
\textbf{SP} & $\mathbf{Y^\nu_{\alpha i}}$ \\
\hline

SP1 &
$\begin{pmatrix}
-0.049 - 0.234 i & -0.123 + 0.006 i & -0.011 - 0.033 i \\
-0.140 + 0.218 i & 0.098 + 0.082 i & -0.018 + 0.037 i \\
-0.052 + 0.279 i & 0.136 + 0.049 i & -0.002 + 0.041 i
\end{pmatrix}$ \\

\hline

SP2 &
$\begin{pmatrix}
-0.017 + 0.005 i & -0.011 - 0.021 i & -0.009 - 0.043 i \\
0.016 + 0.012 i & -0.003 + 0.016 i & -0.025 + 0.044 i \\
0.016 + 0.002 i & 0.011 + 0.018 i & -0.001 + 0.050 i
\end{pmatrix}$ \\

\hline

SP3 &
$\begin{pmatrix}
-0.037 + 0.032 i & -0.013 + 0.026 i & -0.017 - 0.017 i \\
0.001 - 0.062 i & -0.004 - 0.039 i & 0.025 + 0.002 i \\
0.020 - 0.047 i & 0.001 - 0.037 i & 0.024 + 0.012 i
\end{pmatrix}$ \\

\hline

SP4 &
$\begin{pmatrix}
-0.009 + 0.042 i & -0.037 + 0.005 i & 0.009 + 0.004 i \\
0.047 - 0.018 i & 0.042 + 0.030 i & -0.011 - 0.008 i \\
0.022 - 0.031 i & 0.040 + 0.014 i & -0.012 - 0.009 i
\end{pmatrix}$ \\

\hline

SP5 &
$\begin{pmatrix}
-0.016 - 0.024 i & -0.003 + 0.091 i & 0.067 + 0.046 i \\
-0.007 + 0.030 i & 0.073 - 0.069 i & -0.024 - 0.091 i \\
0.007 + 0.034 i & 0.043 - 0.099 i & -0.052 - 0.079 i
\end{pmatrix}$ \\

\hline
\end{tabular}
\caption{Yukawa coupling matrix $Y$ for benchmark points (rounded to three decimal places)}
\label{tab:App-Ymatrix}

\end{table}
\begin{table}[h]
\centering
\renewcommand{\arraystretch}{1.3}
\setlength{\arraycolsep}{8pt}
\begin{tabular}{|c|c|}
\hline
\textbf{SP} & $\mathbf{R}$ \\
\hline

SP1 &
$\begin{pmatrix}
6.699 - 2.178 i & 2.341 + 6.062 i & -0.143 - 2.809 i \\
-13.772 - 6.096 i & 5.715 - 12.447 i & -2.194 + 5.836 i \\
-4.740 + 14.633 i & -13.163 - 4.327 i & 6.465 + 1.919 i
\end{pmatrix}$ \\

\hline

SP2 &
$\begin{pmatrix}
2.169 - 0.485 i & -0.512 - 1.985 i & -0.466 - 0.077 i \\
2.372 + 2.025 i & 2.137 - 2.365 i & 0.777 + 0.321 i \\
1.418 - 2.646 i & -2.792 - 1.446 i & 0.681 - 0.419 i
\end{pmatrix}$ \\

\hline

SP3 &
$\begin{pmatrix}
2.242 - 1.660 i & 1.768 + 2.110 i & -0.243 + 0.035 i \\
-3.715 - 0.893 i & 1.044 - 3.668 i & -0.848 - 0.600 i \\
-0.106 - 3.779 i & 3.833 + 0.026 i & 0.933 - 0.537 i
\end{pmatrix}$ \\

\hline

SP4 &
$\begin{pmatrix}
1.472 - 2.876 i & 3.012 + 1.399 i & 0.114 + 0.164 i \\
3.449 + 2.683 i & -2.770 + 3.286 i & 0.194 - 0.776 i \\
2.524 - 1.989 i & 1.902 + 2.570 i & -1.261 - 0.104 i
\end{pmatrix}$ \\

\hline

SP5 &
$\begin{pmatrix}
2.641 - 1.811 i & -1.901 - 2.520 i & 0.195 - 0.040 i \\
-6.260 + 2.252 i & 1.886 + 6.582 i & -2.656 - 0.634 i \\
2.822 + 6.691 i & 7.034 - 2.446 i & -0.676 + 2.479 i
\end{pmatrix}$ \\

\hline
\end{tabular}
\caption{Casas-Ibarra matrix $R$ for benchmark points (rounded to three decimal places)}
\label{tab:App-Rmatrix}
\end{table}
\begin{table}[h!]
\centering
\renewcommand{\arraystretch}{1.2}
\begin{tabular}{|c|c|c|c|}
\hline
 & $j=1$ & $j=2$ & $j=3$ \\
\hline
$i=1$ & $Y_{11}=0.00297811$ & $Y_{12}=-0.00248201$ & $Y_{13}=-0.00322561$ \\
$i=2$ & $Y_{21}=-0.00248201$ & $Y_{22}=0.00402477$ & $Y_{23}=0.00360735$ \\
$i=3$ & $Y_{31}=-0.00322561$ & $Y_{32}=0.00360735$ & $Y_{33}=0.00409712$ \\
\hline
\end{tabular}
\caption{Numerical values of the Yukawa matrix elements $\mathbf{Y^N_{ij}}$.}
\label{tab:MNyukawa}
\end{table}

\section{Field Dependent Masses}
\label{app:fielddependentmass}

The field-dependent masses are required to compute the one-loop Coleman-Weinberg potential, thermal corrections, and Daisy resummations. These are derived by expanding the scalar fields around classical background values.

We parameterize the fields around their vacuum expectation values (VEVs) as:
\begin{align}
H = \frac{1}{\sqrt{2}} \begin{pmatrix} G^+ \\ h + v_H + i G^0 \end{pmatrix}, \qquad
\Phi = \frac{1}{\sqrt{2}}(h_\Phi + v_\Phi) + i \frac{\chi}{\sqrt{2}},
\label{eq:CP_even_odd}
\end{align}
where \( h \) and \( h_\Phi \) are the CP-even fluctuations, \( \chi \) is the CP-odd scalar, and \( G^0, G^\pm \) are the would-be Goldstone bosons.

To compute the field-dependent masses, we evaluate the second derivatives of the potential with respect to the scalar fields \( h \) and \( h_\Phi \), while setting all other fields (e.g., Goldstones and pseudoscalars) to zero. These masses are functions of the background field values \( h \) and \( h_\Phi \), treated as classical field configurations in the effective potential.

The CP-even scalar mass matrix is obtained from the second derivatives of the tree-level potential:
\begin{align}
\mathcal{M}^2(h, h_\Phi) =
\begin{pmatrix}
- \mu_H^2 + 3 \lambda_H h^2 + \frac{1}{2} \lambda_{H\Phi} h_\Phi^2 & \lambda_{H\Phi} h h_\Phi \\
\lambda_{H\Phi} h h_\Phi & - \mu_\Phi^2 + 3 \lambda_\Phi h_\Phi^2 + \frac{1}{2} \lambda_{H\Phi} h^2
\end{pmatrix},
\end{align}
where the eigenvalues correspond to the field-dependent squared masses of the CP-even scalars \( h_1 \) and \( h_2 \).

The CP-odd scalar \( \chi \) has mass-squared:
\begin{align}
m_\chi^2(h, h_\Phi) = -\mu_\Phi^2 + \lambda_\Phi h_\Phi^2 + \frac{1}{2} \lambda_{H\Phi} h^2.
\end{align}

The Goldstone bosons \( G^0 \) and \( G^\pm \) remain massless at tree level in the Landau gauge. The field-dependent masses of gauge bosons are:
\begin{align}
m_W^2(h) &= \frac{g^2}{4} h^2, \\
m_Z^2(h) &= \frac{g^2 + g'^2}{4} h^2, \\
m_{Z_X}^2(h_\Phi) &= g_X^2 Q_\Phi^2 h_\Phi^2.
\end{align}

If gauge kinetic mixing is present, additional mixing terms between \( Z \) and \( Z_X \) should be included.

For the third-generation fermions, we include the top quark and right-handed neutrinos:
\begin{align}
m_t^2(h) &= \frac{y_t^2}{2} h^2, \\
m_{N_i}^2(h_\Phi) &= \frac{y_{ii}^2}{2} h_\Phi^2.
\end{align}

Here \( y_t \) is the top Yukawa coupling, and \( y_{ii} \) are the Yukawa couplings of the right-handed neutrinos.

\begin{table}[h]
\centering
\begin{tabular}{|c|l|c|}
\hline
Particle & Field-dependent Mass Squared & Degrees of Freedom \\
\hline
\( h_1, h_2 \) & Eigenvalues of \( \mathcal{M}^2(h, h_\Phi) \) & 1 each \\
\( \chi \)     & \( -\mu_\Phi^2 + \lambda_\Phi h_\Phi^2 + \frac{1}{2} \lambda_{H\Phi} h^2 \) & 1 \\
\( W^\pm \)    & \( \frac{g^2}{4} h^2 \) & 6 \\
\( Z \)        & \( \frac{g^2 + g'^2}{4} h^2 \) & 3 \\
\( Z_X \)      & \( g_X^2 Q_\Phi^2 h_\Phi^2 \) & 3 \\
\( t \)        & \( \frac{y_t^2}{2} h^2 \) & $-12$ \\
\( N_i \)      & \( \frac{y_{i}^2}{2} h_\Phi^2 \) & $-2$ per flavor \\
\hline
\end{tabular}
\caption{Field-dependent masses used in one-loop and thermal effective potential. Negative signs for fermionic degrees of freedom follow from the Coleman-Weinberg formalism.}
\end{table}

\section{Derivation of the Counter-term Potential}
\label{app:counterterms}
To ensure the renormalizability of the theory and preserve the vacuum structure at one-loop order, we introduce a counter-term potential that cancels the divergences from the Coleman–Weinberg (CW) correction. This derivation is based on the full scalar potential involving the Standard Model Higgs doublet \( H \) and a complex scalar singlet \( \Phi \).

The scalar potential at tree level is given by:
\begin{align}
V_{\text{tree}}(H, \Phi) = -\mu_H^2 |H|^2 + \lambda_H |H|^4 - \mu_\Phi^2 |\Phi|^2 + \lambda_\Phi |\Phi|^4 + \lambda_{H\Phi} |H|^2 |\Phi|^2.
\label{eq:scalar-potential1}
\end{align}

The scalar fields acquire VEVs as:
\begin{align}
\langle H \rangle = \frac{1}{\sqrt{2}} \begin{pmatrix} 0 \\ v_H \end{pmatrix}, \qquad 
\langle \Phi \rangle = \frac{v_\Phi}{\sqrt{2}}.
\end{align}
    
Substituting Eqn.~\ref{eq:CP_even_odd} into Eqn.~\eqref{eq:scalar-potential1}, the tree-level potential in terms of real scalar fields becomes:
\begin{align}
V_0(h, h_\Phi) &= -\frac{1}{2} \mu_H^2 (h + v_H)^2 + \frac{1}{4} \lambda_H (h + v_H)^4 
- \frac{1}{2} \mu_\Phi^2 (h_\Phi + v_\Phi)^2 + \frac{1}{4} \lambda_\Phi (h_\Phi + v_\Phi)^4 \notag \\
&\quad + \frac{1}{4} \lambda_{H\Phi} (h + v_H)^2 (h_\Phi + v_\Phi)^2.
\end{align}

The zero-temperature one-loop Coleman–Weinberg potential is:
\begin{align}
V_{\text{CW}}(h, h_\Phi) = \sum_i \frac{n_i}{64\pi^2} m_i^4(h, h_\Phi) \left( \log\left( \frac{m_i^2(h, h_\Phi)}{\mu^2} \right) - c_i \right),
\end{align}
where \( m_i(h, h_\Phi) \) are field-dependent masses, \( n_i \) the degrees of freedom, and
\[
c_i = \begin{cases}
\frac{3}{2} & \text{for scalars and fermions}, \\
\frac{5}{6} & \text{for gauge bosons}.
\end{cases}
\]

To cancel divergences and the tree-level vacuum, we introduce a counter-term potential:
\begin{align}
V_{\text{ct}}(h, h_\Phi) 
= -\frac{1}{2} \delta \mu_H^2 h^2 + \frac{1}{4} \delta \lambda_H h^4 
- \frac{1}{2} \delta \mu_\Phi^2 h_\Phi^2 + \frac{1}{4} \delta \lambda_\Phi h_\Phi^4 
+ \frac{1}{4} \delta \lambda_{H\Phi} h^2 h_\Phi^2.
\end{align}

We fix the counterterms by requiring that the first and second derivatives of the effective potential vanish at the tree-level VEVs:
\begin{align}
\left. \frac{\partial V_{\text{CW}}}{\partial h} \right|_{h = v_H,\, h_\Phi = v_\Phi} - \delta \mu_H^2 v_H + \delta \lambda_H v_H^3 + \frac{1}{2}\delta \lambda_{H\Phi} v_H v_\Phi^2 &= 0, \label{vcw1} \\
\left. \frac{\partial V_{\text{CW}}}{\partial h_\Phi} \right|_{h = v_H,\, h_\Phi = v_\Phi} - \delta \mu_\Phi^2 v_\Phi + \delta \lambda_\Phi v_\Phi^3 + \frac{1}{2}\delta \lambda_{H\Phi} v_H^2 v_\Phi &= 0, \label{vcw2}
\end{align}
\begin{align}
\left. \frac{\partial^2 V_{\text{CW}}}{\partial h^2} \right|_{h = v_H,\, h_\Phi = v_\Phi} - \delta \mu_H^2 + 3 \delta \lambda_H v_H^2 + \frac{1}{2}\delta \lambda_{H\Phi} v_\Phi^2 &= 0, \label{vcw3} \\
\left. \frac{\partial^2 V_{\text{CW}}}{\partial h_\Phi^2} \right|_{h = v_H,\, h_\Phi = v_\Phi} - \delta \mu_\Phi^2 + 3 \delta \lambda_\Phi v_\Phi^2 + \frac{1}{2}\delta \lambda_{H\Phi} v_H^2 &= 0, \label{vcw4}\\
\left. \frac{\partial^2 V_{\text{CW}}}{\partial h\, \partial h_\Phi} \right|_{h = v_H,\, h_\Phi = v_\Phi} + \delta \lambda_{H\Phi} v_H v_\Phi &= 0. \label{vcw5}
\end{align}
This system of five equations determines the five counterterms: \( \delta \mu_H^2, \delta \mu_\Phi^2, \delta \lambda_H, \delta \lambda_\Phi, \delta \lambda_{H\Phi} \). With counterterms fixed by conditions \ref{vcw1}-\ref{vcw5} to preserve the physical VEVs and scalar masses at one-loop order.

\bibliographystyle{JHEP}
\bibliography{main}

@article{Esteban:2024eli,
    author = "Esteban, Ivan and Gonzalez-Garcia, M. C. and Maltoni, Michele and Martinez-Soler, Ivan and Pinheiro, Jo{\~a}o Paulo and Schwetz, Thomas",
    title = "{NuFit-6.0: updated global analysis of three-flavor neutrino oscillations}",
    eprint = "2410.05380",
    archivePrefix = "arXiv",
    primaryClass = "hep-ph",
    reportNumber = "IFT-UAM/CSIC-24-140, YITP-SB-2024-24, IPPP/24/64, IPPP/24/64, IFT-UAM/CSIC-24-140, YITP-SB-2024-24",
    doi = "10.1007/JHEP12(2024)216",
    journal = "JHEP",
    volume = "12",
    pages = "216",
    year = "2024"
}

@article{Okada:2025daq,
    author = "Okada, Nobuchika and Raut, Digesh",
    title = "{Analytic formulation of Leptogenesis with neutrino oscillation data employing the general parametrization for neutrino mass matrix}",
    eprint = "2506.20580",
    archivePrefix = "arXiv",
    primaryClass = "hep-ph",
    month = "6",
    year = "2025"
}

@article{Plumacher:1996kc,
    author = "Plumacher, Michael",
    title = "{Baryogenesis and lepton number violation}",
    eprint = "hep-ph/9604229",
    archivePrefix = "arXiv",
    reportNumber = "DESY-96-052",
    doi = "10.1007/s002880050418",
    journal = "Z. Phys. C",
    volume = "74",
    pages = "549--559",
    year = "1997"
}

@article{Wainwright:2011kj,
    author = "Wainwright, Carroll L.",
    title = "{CosmoTransitions: Computing Cosmological Phase Transition Temperatures and Bubble Profiles with Multiple Fields}",
    eprint = "1109.4189",
    archivePrefix = "arXiv",
    primaryClass = "hep-ph",
    doi = "10.1016/j.cpc.2012.04.004",
    journal = "Comput. Phys. Commun.",
    volume = "183",
    pages = "2006--2013",
    year = "2012"
}

@article{Huber:2008hg,
    author = "Huber, Stephan J. and Konstandin, Thomas",
    title = "{Gravitational Wave Production by Collisions: More Bubbles}",
    eprint = "0806.1828",
    archivePrefix = "arXiv",
    primaryClass = "hep-ph",
    doi = "10.1088/1475-7516/2008/09/022",
    journal = "JCAP",
    volume = "09",
    pages = "022",
    year = "2008"
}

@article{Hindmarsh:2013xza,
    author = "Hindmarsh, Mark and Huber, Stephan J. and Rummukainen, Kari and Weir, David J.",
    title = "{Gravitational waves from the sound of a first order phase transition}",
    eprint = "1304.2433",
    archivePrefix = "arXiv",
    primaryClass = "hep-ph",
    reportNumber = "HIP-2013-07-TH",
    doi = "10.1103/PhysRevLett.112.041301",
    journal = "Phys. Rev. Lett.",
    volume = "112",
    pages = "041301",
    year = "2014"
}

@article{Hindmarsh:2015qta,
    author = "Hindmarsh, Mark and Huber, Stephan J. and Rummukainen, Kari and Weir, David J.",
    title = "{Numerical simulations of acoustically generated gravitational waves at a first order phase transition}",
    eprint = "1504.03291",
    archivePrefix = "arXiv",
    primaryClass = "astro-ph.CO",
    reportNumber = "HIP-2015-13-TH",
    doi = "10.1103/PhysRevD.92.123009",
    journal = "Phys. Rev. D",
    volume = "92",
    number = "12",
    pages = "123009",
    year = "2015"
}

@article{Espinosa:2010hh,
    author = "Espinosa, Jose R. and Konstandin, Thomas and No, Jose M. and Servant, Geraldine",
    title = "{Energy Budget of Cosmological First-order Phase Transitions}",
    eprint = "1004.4187",
    archivePrefix = "arXiv",
    primaryClass = "hep-ph",
    reportNumber = "CERN-PH-TH-2010-027",
    doi = "10.1088/1475-7516/2010/06/028",
    journal = "JCAP",
    volume = "06",
    pages = "028",
    year = "2010"
}

@article{Kamionkowski:1993fg,
    author = "Kamionkowski, Marc and Kosowsky, Arthur and Turner, Michael S.",
    title = "{Gravitational radiation from first order phase transitions}",
    eprint = "astro-ph/9310044",
    archivePrefix = "arXiv",
    reportNumber = "IASSNS-HEP-93-44, FERMILAB-PUB-93-235-A",
    doi = "10.1103/PhysRevD.49.2837",
    journal = "Phys. Rev. D",
    volume = "49",
    pages = "2837--2851",
    year = "1994"
}

@article{Kosowsky:1992vn,
    author = "Kosowsky, Arthur and Turner, Michael S.",
    title = "{Gravitational radiation from colliding vacuum bubbles: envelope approximation to many bubble collisions}",
    eprint = "astro-ph/9211004",
    archivePrefix = "arXiv",
    reportNumber = "FERMILAB-PUB-92-295-A",
    doi = "10.1103/PhysRevD.47.4372",
    journal = "Phys. Rev. D",
    volume = "47",
    pages = "4372--4391",
    year = "1993"
}

@article{Ellis:2020nnr,
    author = "Ellis, John and Lewicki, Marek and Vaskonen, Ville",
    title = "{Updated predictions for gravitational waves produced in a strongly supercooled phase transition}",
    eprint = "2007.15586",
    archivePrefix = "arXiv",
    primaryClass = "astro-ph.CO",
    reportNumber = "KCL-PH-TH/2020-40, CERN-TH-2020-129",
    doi = "10.1088/1475-7516/2020/11/020",
    journal = "JCAP",
    volume = "11",
    pages = "020",
    year = "2020"
}

@article{Caprini:2015zlo,
    author = "Caprini, Chiara and others",
    title = "{Science with the space-based interferometer eLISA. II: Gravitational waves from cosmological phase transitions}",
    eprint = "1512.06239",
    archivePrefix = "arXiv",
    primaryClass = "astro-ph.CO",
    reportNumber = "DESY-15-246",
    doi = "10.1088/1475-7516/2016/04/001",
    journal = "JCAP",
    volume = "04",
    pages = "001",
    year = "2016"
}

@article{Guo:2020grp,
    author = "Guo, Huai-Ke and Sinha, Kuver and Vagie, Daniel and White, Graham",
    title = "{Phase Transitions in an Expanding Universe: Stochastic Gravitational Waves in Standard and Non-Standard Histories}",
    eprint = "2007.08537",
    archivePrefix = "arXiv",
    primaryClass = "hep-ph",
    doi = "10.1088/1475-7516/2021/01/001",
    journal = "JCAP",
    volume = "01",
    pages = "001",
    year = "2021"
}

@article{Hindmarsh:2020hop,
    author = {Hindmarsh, Mark B. and L{\"u}ben, Marvin and Lumma, Johannes and Pauly, Martin},
    title = "{Phase transitions in the early universe}",
    eprint = "2008.09136",
    archivePrefix = "arXiv",
    primaryClass = "astro-ph.CO",
    reportNumber = "MPP-2020-163, HIP-2020-27/TH",
    doi = "10.21468/SciPostPhysLectNotes.24",
    journal = "SciPost Phys. Lect. Notes",
    volume = "24",
    pages = "1",
    year = "2021"
}

@article{Caprini:2009yp,
    author = "Caprini, Chiara and Durrer, Ruth and Servant, Geraldine",
    title = "{The stochastic gravitational wave background from turbulence and magnetic fields generated by a first-order phase transition}",
    eprint = "0909.0622",
    archivePrefix = "arXiv",
    primaryClass = "astro-ph.CO",
    doi = "10.1088/1475-7516/2009/12/024",
    journal = "JCAP",
    volume = "12",
    pages = "024",
    year = "2009"
}

@article{Binetruy:2012ze,
    author = "Binetruy, Pierre and Bohe, Alejandro and Caprini, Chiara and Dufaux, Jean-Francois",
    title = "{Cosmological Backgrounds of Gravitational Waves and eLISA/NGO: Phase Transitions, Cosmic Strings and Other Sources}",
    eprint = "1201.0983",
    archivePrefix = "arXiv",
    primaryClass = "gr-qc",
    doi = "10.1088/1475-7516/2012/06/027",
    journal = "JCAP",
    volume = "06",
    pages = "027",
    year = "2012"
}

@article{Banta:2022rwg,
    author = "Banta, Ian",
    title = "{A strongly first-order electroweak phase transition from Loryons}",
    eprint = "2202.04608",
    archivePrefix = "arXiv",
    primaryClass = "hep-ph",
    doi = "10.1007/JHEP06(2022)099",
    journal = "JHEP",
    volume = "06",
    pages = "099",
    year = "2022"
}

@article{Vaskonen:2016yiu,
    author = "Vaskonen, Ville",
    title = "{Electroweak baryogenesis and gravitational waves from a real scalar singlet}",
    eprint = "1611.02073",
    archivePrefix = "arXiv",
    primaryClass = "hep-ph",
    doi = "10.1103/PhysRevD.95.123515",
    journal = "Phys. Rev. D",
    volume = "95",
    number = "12",
    pages = "123515",
    year = "2017"
}

@article{Beniwal:2017eik,
    author = "Beniwal, Ankit and Lewicki, Marek and Wells, James D. and White, Martin and Williams, Anthony G.",
    title = "{Gravitational wave, collider and dark matter signals from a scalar singlet electroweak baryogenesis}",
    eprint = "1702.06124",
    archivePrefix = "arXiv",
    primaryClass = "hep-ph",
    reportNumber = "ADP-17-08-T1014, ADP--17--08-T1014",
    doi = "10.1007/JHEP08(2017)108",
    journal = "JHEP",
    volume = "08",
    pages = "108",
    year = "2017"
}

@article{Kajantie:1996mn,
    author = "Kajantie, K. and Laine, M. and Rummukainen, K. and Shaposhnikov, Mikhail E.",
    title = "{Is there a~ hot electroweak phase transition at $m_H \gtrsim m_W$?}",
    eprint = "hep-ph/9605288",
    archivePrefix = "arXiv",
    reportNumber = "CERN-TH-96-126, HD-THEP-96-15, IUHET-333",
    doi = "10.1103/PhysRevLett.77.2887",
    journal = "Phys. Rev. Lett.",
    volume = "77",
    pages = "2887--2890",
    year = "1996"
}

@article{Su:2020pjw,
    author = "Su, Wei and Williams, Anthony G. and Zhang, Mengchao",
    title = "{Strong first order electroweak phase transition in 2HDM confronting future Z {\&} Higgs factories}",
    eprint = "2011.04540",
    archivePrefix = "arXiv",
    primaryClass = "hep-ph",
    reportNumber = "ADP-20-31/T1141",
    doi = "10.1007/JHEP04(2021)219",
    journal = "JHEP",
    volume = "04",
    pages = "219",
    year = "2021"
}

@article{Ham:2010ha,
    author = "Ham, S. W. and Shim, Seong-A and Oh, S. K.",
    title = "{Electroweak phase transition in an extension of the standard model with scalar color octet}",
    eprint = "1002.0237",
    archivePrefix = "arXiv",
    primaryClass = "hep-ph",
    doi = "10.1103/PhysRevD.81.055015",
    journal = "Phys. Rev. D",
    volume = "81",
    pages = "055015",
    year = "2010"
}

@article{Athron:2019teq,
    author = "Athron, Peter and Balazs, Csaba and Fowlie, Andrew and Pozzo, Giancarlo and White, Graham and Zhang, Yang",
    title = "{Strong first-order phase transitions in the NMSSM {\textemdash} a comprehensive survey}",
    eprint = "1908.11847",
    archivePrefix = "arXiv",
    primaryClass = "hep-ph",
    reportNumber = "CoEPP-MN-19-03",
    doi = "10.1007/JHEP11(2019)151",
    journal = "JHEP",
    volume = "11",
    pages = "151",
    year = "2019"
}

@article{Zhao:2021dxa,
    author = "Zhao, Shu-Min and Zhang, Jian-Fei and Wang, Xi and Dong, Xing-Xing and Feng, Tai-Fu",
    title = "{Strong first-order electroweak phase transition in the U(1)X-extended MSSM}",
    eprint = "2109.07691",
    archivePrefix = "arXiv",
    primaryClass = "hep-ph",
    doi = "10.1103/PhysRevD.108.075031",
    journal = "Phys. Rev. D",
    volume = "108",
    number = "7",
    pages = "075031",
    year = "2023"
}

@article{Coleman:1973jx,
    author = "Coleman, Sidney R. and Weinberg, Erick J.",
    title = "{Radiative Corrections as the Origin of Spontaneous Symmetry Breaking}",
    doi = "10.1103/PhysRevD.7.1888",
    journal = "Phys. Rev. D",
    volume = "7",
    pages = "1888--1910",
    year = "1973"
}

@article{Dorsch:2014qja,
    author = "Dorsch, G. C. and Huber, S. J. and Mimasu, K. and No, J. M.",
    title = "{Echoes of the Electroweak Phase Transition: Discovering a second Higgs doublet through $A_0 \rightarrow ZH_0$}",
    eprint = "1405.5537",
    archivePrefix = "arXiv",
    primaryClass = "hep-ph",
    doi = "10.1103/PhysRevLett.113.211802",
    journal = "Phys. Rev. Lett.",
    volume = "113",
    number = "21",
    pages = "211802",
    year = "2014"
}

@article{Gabelmann:2021ohf,
    author = {Gabelmann, Martin and M{\"u}hlleitner, M. Margarete and M{\"u}ller, Jonas},
    title = "{Electroweak phase transitions with BSM fermions}",
    eprint = "2107.09617",
    archivePrefix = "arXiv",
    primaryClass = "hep-ph",
    reportNumber = "KA-TP-16-2021",
    doi = "10.1007/JHEP01(2022)012",
    journal = "JHEP",
    volume = "01",
    pages = "012",
    year = "2022"
}

@article{Camargo-Molina:2024sde,
    author = {Camargo-Molina, Eliel and Enberg, Rikard and L{\"o}fgren, Johan},
    title = "{A Catalog of First-Order Electroweak Phase Transitions in the Standard Model Effective Field Theory}",
    eprint = "2410.23210",
    archivePrefix = "arXiv",
    primaryClass = "hep-ph",
    month = "10",
    year = "2024"
}

@article{Qin:2024dfp,
    author = "Qin, Renhui and Bian, Ligong",
    title = "{First-order electroweak phase transition with a gauge-invariant approach}",
    eprint = "2408.09677",
    archivePrefix = "arXiv",
    primaryClass = "hep-ph",
    doi = "10.1103/PhysRevD.111.L051702",
    journal = "Phys. Rev. D",
    volume = "111",
    number = "5",
    pages = "L051702",
    year = "2025"
}

@article{Baldes:2016rqn,
    author = "Baldes, Iason and Konstandin, Thomas and Servant, Geraldine",
    title = "{A first-order electroweak phase transition from varying Yukawas}",
    eprint = "1604.04526",
    archivePrefix = "arXiv",
    primaryClass = "hep-ph",
    reportNumber = "DESY-16-068",
    doi = "10.1016/j.physletb.2018.10.015",
    journal = "Phys. Lett. B",
    volume = "786",
    pages = "373--377",
    year = "2018"
}

@article{Harry:2006fi,
    author = "Harry, G. M. and Fritschel, P. and Shaddock, D. A. and Folkner, W. and Phinney, E. S.",
    title = "{Laser interferometry for the big bang observer}",
    doi = "10.1088/0264-9381/23/15/008",
    journal = "Class. Quant. Grav.",
    volume = "23",
    pages = "4887--4894",
    year = "2006",
    note = "[Erratum: Class.Quant.Grav. 23, 7361 (2006)]"
}

@article{Seto:2001qf,
    author = "Seto, Naoki and Kawamura, Seiji and Nakamura, Takashi",
    title = "{Possibility of direct measurement of the acceleration of the universe using 0.1-Hz band laser interferometer gravitational wave antenna in space}",
    eprint = "astro-ph/0108011",
    archivePrefix = "arXiv",
    doi = "10.1103/PhysRevLett.87.221103",
    journal = "Phys. Rev. Lett.",
    volume = "87",
    pages = "221103",
    year = "2001"
}

@article{Harry:2010zz,
    author = "Harry, Gregory M.",
    editor = "Marka, Zsuzsa and Marka, Szabolcs",
    collaboration = "LIGO Scientific",
    title = "{Advanced LIGO: The next generation of gravitational wave detectors}",
    doi = "10.1088/0264-9381/27/8/084006",
    journal = "Class. Quant. Grav.",
    volume = "27",
    pages = "084006",
    year = "2010"
}

@article{Punturo:2010zz,
    author = "Punturo, M. and others",
    editor = "Ricci, Fulvio",
    title = "{The Einstein Telescope: A third-generation gravitational wave observatory}",
    doi = "10.1088/0264-9381/27/19/194002",
    journal = "Class. Quant. Grav.",
    volume = "27",
    pages = "194002",
    year = "2010"
}

@article{Manohar:2006ga,
    author = "Manohar, Aneesh V. and Wise, Mark B.",
    title = "{Flavor changing neutral currents, an extended scalar sector, and the Higgs production rate at the CERN LHC}",
    eprint = "hep-ph/0606172",
    archivePrefix = "arXiv",
    reportNumber = "UCSD-PTH-06-07, CALT-68-2601",
    doi = "10.1103/PhysRevD.74.035009",
    journal = "Phys. Rev. D",
    volume = "74",
    pages = "035009",
    year = "2006"
}

@article{Profumo:2007wc,
    author = "Profumo, Stefano and Ramsey-Musolf, Michael J. and Shaughnessy, Gabe",
    title = "{Singlet Higgs phenomenology and the electroweak phase transition}",
    eprint = "0705.2425",
    archivePrefix = "arXiv",
    primaryClass = "hep-ph",
    reportNumber = "CALTECH-MAP-333, MADPH-07-1489",
    doi = "10.1088/1126-6708/2007/08/010",
    journal = "JHEP",
    volume = "08",
    pages = "010",
    year = "2007"
}

@article{LIGOScientific:2016aoc,
    author = "Abbott, B. P. and others",
    collaboration = "LIGO Scientific, Virgo",
    title = "{Observation of Gravitational Waves from a Binary Black Hole Merger}",
    eprint = "1602.03837",
    archivePrefix = "arXiv",
    primaryClass = "gr-qc",
    reportNumber = "LIGO-P150914",
    doi = "10.1103/PhysRevLett.116.061102",
    journal = "Phys. Rev. Lett.",
    volume = "116",
    number = "6",
    pages = "061102",
    year = "2016"
}

@article{Minkowski:1977sc,
    author = "Minkowski, Peter",
    title = "{$\mu \to e\gamma$ at a Rate of One Out of $10^{9}$ Muon Decays?}",
    reportNumber = "Print-77-0182 (BERN)",
    doi = "10.1016/0370-2693(77)90435-X",
    journal = "Phys. Lett. B",
    volume = "67",
    pages = "421--428",
    year = "1977"
}

@article{Yanagida:1980xy,
    author = "Yanagida, Tsutomu",
    title = "{Horizontal Symmetry and Masses of Neutrinos}",
    reportNumber = "TU-80-208",
    doi = "10.1143/PTP.64.1103",
    journal = "Prog. Theor. Phys.",
    volume = "64",
    pages = "1103",
    year = "1980"
}

@article{Yanagida:1979as,
    author = "Yanagida, Tsutomu",
    editor = "Sawada, Osamu and Sugamoto, Akio",
    title = "{Horizontal gauge symmetry and masses of neutrinos}",
    reportNumber = "KEK-79-18-95",
    journal = "Conf. Proc. C",
    volume = "7902131",
    pages = "95--99",
    year = "1979"
}

@article{Gell-Mann:1979vob,
    author = "Gell-Mann, Murray and Ramond, Pierre and Slansky, Richard",
    title = "{Complex Spinors and Unified Theories}",
    eprint = "1306.4669",
    archivePrefix = "arXiv",
    primaryClass = "hep-th",
    reportNumber = "PRINT-80-0576",
    journal = "Conf. Proc. C",
    volume = "790927",
    pages = "315--321",
    year = "1979"
}

@article{Mohapatra:1979ia,
    author = "Mohapatra, Rabindra N. and Senjanovic, Goran",
    title = "{Neutrino Mass and Spontaneous Parity Nonconservation}",
    reportNumber = "MDDP-TR-80-060, MDDP-PP-80-105, CCNY-HEP-79-10",
    doi = "10.1103/PhysRevLett.44.912",
    journal = "Phys. Rev. Lett.",
    volume = "44",
    pages = "912",
    year = "1980"
}

@article{Schmitz:2020syl,
    author = "Schmitz, Kai",
    title = "{New Sensitivity Curves for Gravitational-Wave Signals from Cosmological Phase Transitions}",
    eprint = "2002.04615",
    archivePrefix = "arXiv",
    primaryClass = "hep-ph",
    reportNumber = "CERN-TH-2020-018",
    doi = "10.1007/JHEP01(2021)097",
    journal = "JHEP",
    volume = "01",
    pages = "097",
    year = "2021"
}

@article{Caprini:2018mtu,
    author = "Caprini, Chiara and Figueroa, Daniel G.",
    title = "{Cosmological Backgrounds of Gravitational Waves}",
    eprint = "1801.04268",
    archivePrefix = "arXiv",
    primaryClass = "astro-ph.CO",
    doi = "10.1088/1361-6382/aac608",
    journal = "Class. Quant. Grav.",
    volume = "35",
    number = "16",
    pages = "163001",
    year = "2018"
}

@article{Mazumdar:2018dfl,
    author = "Mazumdar, Anupam and White, Graham",
    title = "{Review of cosmic phase transitions: their significance and experimental signatures}",
    eprint = "1811.01948",
    archivePrefix = "arXiv",
    primaryClass = "hep-ph",
    doi = "10.1088/1361-6633/ab1f55",
    journal = "Rept. Prog. Phys.",
    volume = "82",
    number = "7",
    pages = "076901",
    year = "2019"
}

@article{Caprini:2019egz,
    author = "Caprini, Chiara and others",
    title = "{Detecting gravitational waves from cosmological phase transitions with LISA: an update}",
    eprint = "1910.13125",
    archivePrefix = "arXiv",
    primaryClass = "astro-ph.CO",
    reportNumber = "DESY-19-159, IPPP/19/27, HIP-2019-14/TH, MITP/19-066, IFT-UAM/CSIC-19-139",
    doi = "10.1088/1475-7516/2020/03/024",
    journal = "JCAP",
    volume = "03",
    pages = "024",
    year = "2020"
}

@article{amaro2017laser,
  title={Laser interferometer space antenna},
  author={Amaro-Seoane, Pau and Audley, Heather and Babak, Stanislav and Baker, John and Barausse, Enrico and Bender, Peter and Berti, Emanuele and Binetruy, Pierre and Born, Michael and Bortoluzzi, Daniele and others},
  journal={arXiv preprint arXiv:1702.00786},
  year={2017}
}

@article{Kawamura:2011zz,
    author = "Kawamura, Seiji and others",
    editor = "Buchman, Sasha and Sun, Ke-Xun",
    title = "{The Japanese space gravitational wave antenna: DECIGO}",
    doi = "10.1088/0264-9381/28/9/094011",
    journal = "Class. Quant. Grav.",
    volume = "28",
    pages = "094011",
    year = "2011"
}

@article{NANOGrav:2020bcs,
    author = "Arzoumanian, Zaven and others",
    collaboration = "NANOGrav",
    title = "{The NANOGrav 12.5 yr Data Set: Search for an Isotropic Stochastic Gravitational-wave Background}",
    eprint = "2009.04496",
    archivePrefix = "arXiv",
    primaryClass = "astro-ph.HE",
    doi = "10.3847/2041-8213/abd401",
    journal = "Astrophys. J. Lett.",
    volume = "905",
    number = "2",
    pages = "L34",
    year = "2020"
}

@article{Addazi:2020zcj,
    author = "Addazi, Andrea and Cai, Yi-Fu and Gan, Qingyu and Marciano, Antonino and Zeng, Kaiqiang",
    title = "{NANOGrav results and dark first order phase transitions}",
    eprint = "2009.10327",
    archivePrefix = "arXiv",
    primaryClass = "hep-ph",
    doi = "10.1007/s11433-021-1724-6",
    journal = "Sci. China Phys. Mech. Astron.",
    volume = "64",
    number = "9",
    pages = "290411",
    year = "2021"
}

@article{Nakai:2020oit,
    author = "Nakai, Yuichiro and Suzuki, Motoo and Takahashi, Fuminobu and Yamada, Masaki",
    title = "{Gravitational Waves and Dark Radiation from Dark Phase Transition: Connecting NANOGrav Pulsar Timing Data and Hubble Tension}",
    eprint = "2009.09754",
    archivePrefix = "arXiv",
    primaryClass = "astro-ph.CO",
    reportNumber = "TU-1109; IPMU20-0100",
    doi = "10.1016/j.physletb.2021.136238",
    journal = "Phys. Lett. B",
    volume = "816",
    pages = "136238",
    year = "2021"
}

@article{Li:2020cjj,
    author = "Li, Hao-Hao and Ye, Gen and Piao, Yun-Song",
    title = "{Is the NANOGrav signal a hint of dS decay during inflation?}",
    eprint = "2009.14663",
    archivePrefix = "arXiv",
    primaryClass = "astro-ph.CO",
    doi = "10.1016/j.physletb.2021.136211",
    journal = "Phys. Lett. B",
    volume = "816",
    pages = "136211",
    year = "2021"
}

@article{Ratzinger:2020koh,
    author = "Ratzinger, Wolfram and Schwaller, Pedro",
    title = "{Whispers from the dark side: Confronting light new physics with NANOGrav data}",
    eprint = "2009.11875",
    archivePrefix = "arXiv",
    primaryClass = "astro-ph.CO",
    reportNumber = "MITP/20-056",
    doi = "10.21468/SciPostPhys.10.2.047",
    journal = "SciPost Phys.",
    volume = "10",
    number = "2",
    pages = "047",
    year = "2021"
}

@book{Peskin:1995ev,
    author = "Peskin, Michael E. and Schroeder, Daniel V.",
    title = "{An Introduction to quantum field theory}",
    doi = "10.1201/9780429503559",
    isbn = "978-0-201-50397-5, 978-0-429-50355-9, 978-0-429-49417-8",
    publisher = "Addison-Wesley",
    address = "Reading, USA",
    year = "1995"
}

@article{Delaunay:2007wb,
    author = "Delaunay, Cedric and Grojean, Christophe and Wells, James D.",
    title = "{Dynamics of Non-renormalizable Electroweak Symmetry Breaking}",
    eprint = "0711.2511",
    archivePrefix = "arXiv",
    primaryClass = "hep-ph",
    reportNumber = "CERN-PH-TH-2007-219, MCTP-07-31, SACLAY-T07-141",
    doi = "10.1088/1126-6708/2008/04/029",
    journal = "JHEP",
    volume = "04",
    pages = "029",
    year = "2008"
}

@article{Espinosa:2007qk,
    author = "Espinosa, Jose Ramon and Quiros, Mariano",
    title = "{Novel Effects in Electroweak Breaking from a Hidden Sector}",
    eprint = "hep-ph/0701145",
    archivePrefix = "arXiv",
    reportNumber = "IFT-UAM-CSIC-07-01, UAB-FT-623",
    doi = "10.1103/PhysRevD.76.076004",
    journal = "Phys. Rev. D",
    volume = "76",
    pages = "076004",
    year = "2007"
}

@article{Noble:2007kk,
    author = "Noble, Andrew and Perelstein, Maxim",
    title = "{Higgs self-coupling as a probe of electroweak phase transition}",
    eprint = "0711.3018",
    archivePrefix = "arXiv",
    primaryClass = "hep-ph",
    doi = "10.1103/PhysRevD.78.063518",
    journal = "Phys. Rev. D",
    volume = "78",
    pages = "063518",
    year = "2008"
}

@article{Espinosa:2008kw,
    author = "Espinosa, J. R. and Konstandin, T. and No, J. M. and Quiros, M.",
    title = "{Some Cosmological Implications of Hidden Sectors}",
    eprint = "0809.3215",
    archivePrefix = "arXiv",
    primaryClass = "hep-ph",
    reportNumber = "CERN-PH-TH-2008-196, IFT-UAM-CSIC-08-54, UAB-FT-655",
    doi = "10.1103/PhysRevD.78.123528",
    journal = "Phys. Rev. D",
    volume = "78",
    pages = "123528",
    year = "2008"
}

@article{Espinosa:2011ax,
    author = "Espinosa, Jose R. and Konstandin, Thomas and Riva, Francesco",
    title = "{Strong Electroweak Phase Transitions in the Standard Model with a Singlet}",
    eprint = "1107.5441",
    archivePrefix = "arXiv",
    primaryClass = "hep-ph",
    reportNumber = "CERN-PH-TH-2011-171",
    doi = "10.1016/j.nuclphysb.2011.09.010",
    journal = "Nucl. Phys. B",
    volume = "854",
    pages = "592--630",
    year = "2012"
}

@article{Curtin:2014jma,
    author = "Curtin, David and Meade, Patrick and Yu, Chiu-Tien",
    title = "{Testing Electroweak Baryogenesis with Future Colliders}",
    eprint = "1409.0005",
    archivePrefix = "arXiv",
    primaryClass = "hep-ph",
    reportNumber = "YITP-SB-14-33",
    doi = "10.1007/JHEP11(2014)127",
    journal = "JHEP",
    volume = "11",
    pages = "127",
    year = "2014"
}

@article{Blinov:2015sna,
    author = "Blinov, Nikita and Kozaczuk, Jonathan and Morrissey, David E. and Tamarit, Carlos",
    title = "{Electroweak Baryogenesis from Exotic Electroweak Symmetry Breaking}",
    eprint = "1504.05195",
    archivePrefix = "arXiv",
    primaryClass = "hep-ph",
    reportNumber = "IPPP-15-23, DCPT-15-46",
    doi = "10.1103/PhysRevD.92.035012",
    journal = "Phys. Rev. D",
    volume = "92",
    number = "3",
    pages = "035012",
    year = "2015"
}

@article{Basler:2016obg,
    author = "Basler, P. and Krause, M. and Muhlleitner, M. and Wittbrodt, J. and Wlotzka, A.",
    title = "{Strong First Order Electroweak Phase Transition in the CP-Conserving 2HDM Revisited}",
    eprint = "1612.04086",
    archivePrefix = "arXiv",
    primaryClass = "hep-ph",
    doi = "10.1007/JHEP02(2017)121",
    journal = "JHEP",
    volume = "02",
    pages = "121",
    year = "2017"
}

@article{Basler:2017uxn,
    author = {Basler, Philipp and M{\"u}hlleitner, Margarete and Wittbrodt, Jonas},
    title = "{The CP-Violating 2HDM in Light of a Strong First Order Electroweak Phase Transition and Implications for Higgs Pair Production}",
    eprint = "1711.04097",
    archivePrefix = "arXiv",
    primaryClass = "hep-ph",
    reportNumber = "DESY-17-174, KA-TP-39-2017",
    doi = "10.1007/JHEP03(2018)061",
    journal = "JHEP",
    volume = "03",
    pages = "061",
    year = "2018"
}

@article{Chala:2018ari,
    author = "Chala, Mikael and Krause, Claudius and Nardini, Germano",
    title = "{Signals of the electroweak phase transition at colliders and gravitational wave observatories}",
    eprint = "1802.02168",
    archivePrefix = "arXiv",
    primaryClass = "hep-ph",
    reportNumber = "FERMILAB-PUB-18-241-T",
    doi = "10.1007/JHEP07(2018)062",
    journal = "JHEP",
    volume = "07",
    pages = "062",
    year = "2018"
}

@article{Turner:1990rc,
    author = "Turner, Michael S. and Wilczek, Frank",
    title = "{Relic gravitational waves and extended inflation}",
    reportNumber = "FERMILAB-PUB-90-178-A",
    doi = "10.1103/PhysRevLett.65.3080",
    journal = "Phys. Rev. Lett.",
    volume = "65",
    pages = "3080--3083",
    year = "1990"
}

@article{Kosowsky:1991ua,
    author = "Kosowsky, Arthur and Turner, Michael S. and Watkins, Richard",
    title = "{Gravitational radiation from colliding vacuum bubbles}",
    reportNumber = "FERMILAB-PUB-91-323-A",
    doi = "10.1103/PhysRevD.45.4514",
    journal = "Phys. Rev. D",
    volume = "45",
    pages = "4514--4535",
    year = "1992"
}

@article{Kosowsky:1992rz,
    author = "Kosowsky, Arthur and Turner, Michael S. and Watkins, Richard",
    title = "{Gravitational waves from first order cosmological phase transitions}",
    reportNumber = "FERMILAB-PUB-91-333-A-REV, FERMILAB-PUB-91-333-A",
    doi = "10.1103/PhysRevLett.69.2026",
    journal = "Phys. Rev. Lett.",
    volume = "69",
    pages = "2026--2029",
    year = "1992"
}

@article{Turner:1992tz,
    author = "Turner, Michael S. and Weinberg, Erick J. and Widrow, Lawrence M.",
    title = "{Bubble nucleation in first order inflation and other cosmological phase transitions}",
    reportNumber = "FERMILAB-PUB-91-334-A, CU-TP-558, IASSNS-HEP-92-21",
    doi = "10.1103/PhysRevD.46.2384",
    journal = "Phys. Rev. D",
    volume = "46",
    pages = "2384--2403",
    year = "1992"
}

@article{Kosowsky:2001xp,
    author = "Kosowsky, Arthur and Mack, Andrew and Kahniashvili, Tinatin",
    title = "{Gravitational radiation from cosmological turbulence}",
    eprint = "astro-ph/0111483",
    archivePrefix = "arXiv",
    reportNumber = "RAP-334",
    doi = "10.1103/PhysRevD.66.024030",
    journal = "Phys. Rev. D",
    volume = "66",
    pages = "024030",
    year = "2002"
}

@article{Dolgov:2002ra,
    author = "Dolgov, Alexander D. and Grasso, Dario and Nicolis, Alberto",
    title = "{Relic backgrounds of gravitational waves from cosmic turbulence}",
    eprint = "astro-ph/0206461",
    archivePrefix = "arXiv",
    doi = "10.1103/PhysRevD.66.103505",
    journal = "Phys. Rev. D",
    volume = "66",
    pages = "103505",
    year = "2002"
}

@article{Gogoberidze:2007an,
    author = "Gogoberidze, Grigol and Kahniashvili, Tina and Kosowsky, Arthur",
    title = "{The Spectrum of Gravitational Radiation from Primordial Turbulence}",
    eprint = "0705.1733",
    archivePrefix = "arXiv",
    primaryClass = "astro-ph",
    doi = "10.1103/PhysRevD.76.083002",
    journal = "Phys. Rev. D",
    volume = "76",
    pages = "083002",
    year = "2007"
}

@article{Hindmarsh:2016lnk,
    author = "Hindmarsh, Mark",
    title = "{Sound shell model for acoustic gravitational wave production at a first-order phase transition in the early Universe}",
    eprint = "1608.04735",
    archivePrefix = "arXiv",
    primaryClass = "astro-ph.CO",
    doi = "10.1103/PhysRevLett.120.071301",
    journal = "Phys. Rev. Lett.",
    volume = "120",
    number = "7",
    pages = "071301",
    year = "2018"
}

@book{Bellac:2011kqa,
    author = "Bellac, Michel Le",
    title = "{Thermal Field Theory}",
    doi = "10.1017/CBO9780511721700",
    isbn = "978-0-511-88506-8, 978-0-521-65477-7",
    publisher = "Cambridge University Press",
    series = "Cambridge Monographs on Mathematical Physics",
    month = "3",
    year = "2011"
}

@book{Kapusta:2006pm,
    author = "Kapusta, J. I. and Gale, Charles",
    title = "{Finite-temperature field theory: Principles and applications}",
    doi = "10.1017/CBO9780511535130",
    isbn = "978-0-521-17322-3, 978-0-521-82082-0, 978-0-511-22280-1",
    publisher = "Cambridge University Press",
    series = "Cambridge Monographs on Mathematical Physics",
    year = "2011"
}

@article{Davidson:2002qv,
    author = "Davidson, Sacha and Ibarra, Alejandro",
    title = "{A Lower bound on the right-handed neutrino mass from leptogenesis}",
    eprint = "hep-ph/0202239",
    archivePrefix = "arXiv",
    reportNumber = "OUTP-02-10P, IPPP-02-16, DCPT-02-32",
    doi = "10.1016/S0370-2693(02)01735-5",
    journal = "Phys. Lett. B",
    volume = "535",
    pages = "25--32",
    year = "2002"
}

@article{Sakharov:1967dj,
    author = "Sakharov, A. D.",
    title = "{Violation of CP Invariance, C asymmetry, and baryon asymmetry of the universe}",
    doi = "10.1070/PU1991v034n05ABEH002497",
    journal = "Pisma Zh. Eksp. Teor. Fiz.",
    volume = "5",
    pages = "32--35",
    year = "1967"
}

@article{Okada:2020vvb,
    author = "Okada, Nobuchika and Seto, Osamu and Uchida, Hikaru",
    title = "{Gravitational waves from breaking of an extra $U(1)$ in $SO(10)$ grand unification}",
    eprint = "2006.01406",
    archivePrefix = "arXiv",
    primaryClass = "hep-ph",
    reportNumber = "EPHOU-20-006",
    doi = "10.1093/ptep/ptab003",
    journal = "PTEP",
    volume = "2021",
    number = "3",
    pages = "033B01",
    year = "2021"
}

@article{Okada:2018xdh,
    author = "Okada, Nobuchika and Seto, Osamu",
    title = "{Probing the seesaw scale with gravitational waves}",
    eprint = "1807.00336",
    archivePrefix = "arXiv",
    primaryClass = "hep-ph",
    reportNumber = "EPHOU-18-007",
    doi = "10.1103/PhysRevD.98.063532",
    journal = "Phys. Rev. D",
    volume = "98",
    number = "6",
    pages = "063532",
    year = "2018"
}

@article{Fu:2023nrn,
    author = "Fu, Bowen and Ghoshal, Anish and King, Stephen F.",
    title = "{Cosmic string gravitational waves from global U(1)$_{B−L}$ symmetry breaking as a probe of the type I seesaw scale}",
    eprint = "2306.07334",
    archivePrefix = "arXiv",
    primaryClass = "hep-ph",
    doi = "10.1007/JHEP11(2023)071",
    journal = "JHEP",
    volume = "11",
    pages = "071",
    year = "2023"
}

@article{Chaudhuri:2025ybh,
    author = "Chaudhuri, Arnab",
    title = "{Gravitational Waves from First-Order Phase Transitions Assisted by Temperature-Enhanced Scatterings}",
    eprint = "2507.13135",
    archivePrefix = "arXiv",
    primaryClass = "astro-ph.CO",
    month = "7",
    year = "2025"
}

@article{Srivastava:2025oer,
    author = "Srivastava, Tripurari and Das, Jaydeb and Ghosh, Anupam and Chaudhuri, Arnab",
    title = "{Electroweak Phase Transition, Gravitational Waves and Collider Probes in Multi-Scalar Dark Matter Scenarios}",
    eprint = "2507.05917",
    archivePrefix = "arXiv",
    primaryClass = "hep-ph",
    month = "7",
    year = "2025"
}

@article{Chaudhuri:2024vrd,
    author = "Chaudhuri, Arnab and Kohri, Kazunori",
    title = "{The N2HDM, Entropy Production and Stochastic Gravitational Waves}",
    eprint = "2404.10288",
    archivePrefix = "arXiv",
    primaryClass = "hep-ph",
    month = "4",
    year = "2024"
}

@article{Haba:2019qol,
    author = "Haba, Naoyuki and Yamada, Toshifumi",
    title = "{Gravitational waves from phase transition in minimal SUSY $U(1)_{B-L}$  model}",
    eprint = "1911.01292",
    archivePrefix = "arXiv",
    primaryClass = "hep-ph",
    doi = "10.1103/PhysRevD.101.075027",
    journal = "Phys. Rev. D",
    volume = "101",
    number = "7",
    pages = "075027",
    year = "2020"
}

@article{Brdar:2018num,
    author = "Brdar, Vedran and Helmboldt, Alexander J. and Kubo, Jisuke",
    title = "{Gravitational Waves from First-Order Phase Transitions: LIGO as a Window to Unexplored Seesaw Scales}",
    eprint = "1810.12306",
    archivePrefix = "arXiv",
    primaryClass = "hep-ph",
    doi = "10.1088/1475-7516/2019/02/021",
    journal = "JCAP",
    volume = "02",
    pages = "021",
    year = "2019"
}

@article{Bandyopadhyay:2021ipw,
    author = "Bandyopadhyay, Priyotosh and Jangid, Shilpa",
    title = "{Discerning singlet and triplet scalars at the electroweak phase transition and gravitational wave}",
    eprint = "2111.03866",
    archivePrefix = "arXiv",
    primaryClass = "hep-ph",
    reportNumber = "IITH-PH- 0003-21",
    doi = "10.1103/PhysRevD.107.055032",
    journal = "Phys. Rev. D",
    volume = "107",
    number = "5",
    pages = "055032",
    year = "2023"
}

@article{Oda:2015gna,
    author = "Oda, Satsuki and Okada, Nobuchika and Takahashi, Dai-suke",
    title = "{Classically conformal U(1)' extended standard model and Higgs vacuum stability}",
    eprint = "1504.06291",
    archivePrefix = "arXiv",
    primaryClass = "hep-ph",
    doi = "10.1103/PhysRevD.92.015026",
    journal = "Phys. Rev. D",
    volume = "92",
    number = "1",
    pages = "015026",
    year = "2015"
}

@article{Das:2016zue,
    author = "Das, Arindam and Oda, Satsuki and Okada, Nobuchika and Takahashi, Dai-suke",
    title = "{Classically conformal U(1)' extended standard model, electroweak vacuum stability, and LHC Run-2 bounds}",
    eprint = "1605.01157",
    archivePrefix = "arXiv",
    primaryClass = "hep-ph",
    doi = "10.1103/PhysRevD.93.115038",
    journal = "Phys. Rev. D",
    volume = "93",
    number = "11",
    pages = "115038",
    year = "2016"
}

@article{Gola:2022nkg,
    author = "Gola, Shivam",
    title = "{Pseudo scalar dark matter in a generic U(1)X model}",
    eprint = "2212.04698",
    archivePrefix = "arXiv",
    primaryClass = "hep-ph",
    doi = "10.1016/j.physletb.2023.137982",
    journal = "Phys. Lett. B",
    volume = "842",
    pages = "137982",
    year = "2023"
}

@article{Hasegawa:2019amx,
    author = "Hasegawa, Taiki and Okada, Nobuchika and Seto, Osamu",
    title = "{Gravitational waves from the minimal gauged $U(1)_{B-L}$ model}",
    eprint = "1904.03020",
    archivePrefix = "arXiv",
    primaryClass = "hep-ph",
    reportNumber = "EPHOU-19-004",
    doi = "10.1103/PhysRevD.99.095039",
    journal = "Phys. Rev. D",
    volume = "99",
    number = "9",
    pages = "095039",
    year = "2019"
}

@article{Casas:2001sr,
    author = "Casas, J. A. and Ibarra, A.",
    title = "{Oscillating neutrinos and $\mu \to e, \gamma$}",
    eprint = "hep-ph/0103065",
    archivePrefix = "arXiv",
    reportNumber = "IEM-FT-211-01, OUTP-01-11P, IFT-UAM-CSIC-01-08",
    doi = "10.1016/S0550-3213(01)00475-8",
    journal = "Nucl. Phys. B",
    volume = "618",
    pages = "171--204",
    year = "2001"
}
\end{document}